\documentclass[pra,twocolumn,showpacs,preprintnumbers,superscriptaddress]{revtex4}

\usepackage{amsmath,amsfonts,amssymb,bbm,slashed}
\usepackage[english]{babel}
\usepackage{graphicx}
\usepackage{color}
\usepackage[utf8]{inputenc}
\usepackage[T1]{fontenc}
\usepackage{tensor}

\newcommand{\tgamma}{\tilde{\gamma}}
\newcommand{\tc}{\tilde{c}}

\begin{document}

\title{Dirac equation in 2-dimensional curved spacetime, particle creation, and coupled waveguide arrays}

\author{Christian Koke}
\email{christian.koke@stud.uni-heidelberg.de}
\affiliation{Institut f\"ur theoretische Physik, Philosophenweg 16, D-69120 Heidelberg, Germany.}

\author{Changsuk Noh}
\email{changsuk@kias.re.kr}
\affiliation{Korea Institute for Advanced Study, 85 Hoegiro, Seoul 130-722, Korea.}

\author{Dimitris G. Angelakis}
\email{dimitris.angelakis@gmail.com}
\affiliation{Centre for Quantum Technologies, National University of Singapore, 2 Science Drive 3, Singapore 117542.}
\affiliation{School of Electrical and Computer Engineering, Technical University of Crete, Chania, Crete, Greece, 73100.}

\begin{abstract}
When quantum fields are coupled to gravitational fields, spontaneous particle creation may occur similarly to when they are coupled to external electromagnetic fields. A gravitational field can be incorporated as a background spacetime if the back-action of matter on it can be neglected, resulting in modifications of the Dirac or Klein-Gordon equations for elementary fermions and bosons respectively. The semi-classical description predicts particle creation in many situations, including the expanding-universe scenario, near the event horizon of a black hole (the Hawking effect), and an accelerating observer in flat spacetime (the Unruh effect).  In this work, we give a pedagogical introduction to the Dirac equation in a general 2D spacetime and show examples of spinor wave packet dynamics in flat and curved background spacetimes. In particular, we cover the phenomenon of particle creation in a time-dependent metric. Photonic analogs of these effects are then proposed, where classical light propagating in an array of coupled waveguides provides a visualisation of the Dirac spinor propagating in a curved 2D spacetime background. The extent to which such a single-particle description can be said to mimic particle creation is discussed.
\end{abstract}


\maketitle

\section{Introduction}
Quantum theory of gravity is one of the most sought-after goals in  physics. Despite continuous efforts to tackle this important problem, resulting in  interesting proposals such as superstring theory and loop quantum gravity, there is still no clear sign of the theory of quantum gravity \cite{Giulini,Rovelli,Kiefer}. However, when the back-action of matter on the gravitation field is neglected, one can write down a theory of quantum fields in a background curved spacetime by extending quantum field theory in Minkowski metric to a general metric \cite{Birrell, Fulling, Mukhanov, Parker}. This is analogous to treating external fields as c-numbers and predicts interesting new phenomena that are valid in appropriate regimes. Similarly to the prediction of pair creation in external electric fields \cite{Sauter, HeisenbergEuler1936, Schwinger}, external gravitational fields (as described by a background spacetime) induce particle creation \cite{Parker1971, Hawking1975, Unruh1976}. In the latter, particle creation is caused by the change in the vacuum state itself under quite generic conditions. 

In this work, we propose a classical optical simulation of particle creation in binary waveguide arrays. There have been many proposals and experimental demonstrations of a plethora of interesting physics in coupled waveguide arrays \cite{Peschel, Morandotti, Lahini, Longhi11, Longhi12, Crespi, Keil, Rodriguez13, Rodriguez14, LeeAngelakis14, Marini14, RaiAngelakis15, Keil15}. In particular, optical simulation of the 1+1 dimensional Dirac equation in binary waveguide arrays has been proposed \cite{Longhi10a,Longhi10b} and experimentally demonstrated \cite{Dreisow10,Dreisow12}.  We show that the setup can be generalised to also simulate the Dirac equation in 2 dimensional curved spacetime.

Particle creation is by definition a multi-particle phenomenon and the full simulation of the result requires quantum fields as a main ingredient (for example, see \cite{Boada} for a simulation of the Dirac equation in curved spacetime with cold atoms on optical lattices). However, light propagation in a waveguide array is an intrinsically classical phenomena, so how can we simulate particle creation in a binary waveguide array? The short answer is that we will be looking at a single-particle analog of particle creation. As we will see, the fundamental reason behind particle creation is the difference in the vacuum state, which in turn is captured by different mode-expansions of quantum fields. We can thus concentrate on a single-mode at a time and simulate the effect. In fact, the well-known Klein paradox shows that the single-particle Dirac equation contains subtle hints of multi-particle effects, and the phenomenon of pair production in strong electric fields has been studied within the single particle picture \cite{Ruf}. We study the time evolution of spinor wave packets and demonstrate that an analog of particle creation can be {\it visualised} in the light evolution in a binary waveguide array. Here, we stress that by using the phenomenon of `zitterbewegung' (the jittering motion of a Dirac particle), one can bypass the quantitative checks in `proving' the simulation of particle creation.

This article is organised as follows. In section II, we provide a pedagogical introduction to the Dirac equation in curved spacetime, assuming familiarity with the conventional Dirac equation. In section III, we specialise to the 1+1 dimensions and provide a few examples of spacetime metrics. Particle creation in curved spacetime is explained, using scalar fields for simplicity, and the single-particle analogs for the Dirac spinors is discussed. Section IV shows the wave packet evolution both in flat and curved spacetimes, using time-dependent gravitational fields as an example. A single-particle analog of particle creation in a particular case is explicitly demonstrated. Section V explains the optical simulation of the Dirac equation in a binary waveguide array and proposes a generalisation to curved spacetimes. We conclude in section VI. 

We have tried to be as pedagogical and self-contained as we could. We have tried to collect and present essential ideas to understand quantum fields in curved spacetime and how it predicts particle creation. It is our hope that the reader will find this helpful in understanding the essential ideas quickly and develop further interesting analogies.

\section{Dirac equation in curved spacetime}
Let us start with the derivation of the Dirac equation in curved spacetime. This requires the notion of spin-connection, which will be discussed at a pedagogical level. We closely follow ref.~\cite{Lawrie}.

\subsection{General covariance}
Special theory of relativity has taught us that time and space are observer-dependent concepts in that observers moving relative to each other have different notions of time and space intervals. This comes about because the laws of physics are covariant under a Lorentz transformation, meaning that all observers agree on the form of physical laws in their own coordinate frames. General theory of relativity takes this one step further and states that the laws should be covariant under general coordinate transformations. Equations of motion are written in terms of tensors--quantities that are independent of local coordinate systems used to describe them. To write down the equations of motion one also requires the notion of a covariant derivative, which basically is the correct way of differentiating tensors to yield another tensor (of higher rank). To understand this, consider a vector in Minkowski spacetime $V^\mu$, which transforms under the Lorentz transformation as $V^{\mu'}  = \Lambda \indices{^{\mu' }_{\mu}} V^\mu$. We use the Einstein convention where the repeated indices are summed over. Now let's see how the partial derivative of a vector transforms in flat spacetime:
\begin{align}
\partial_{\nu '}V^{\mu '} = \Lambda \indices{_{\nu '}^{\nu}}\partial_{\nu} \left( \Lambda \indices{^{\mu '}_{\mu}} V^\mu \right) =  \Lambda \indices{_{\nu '}^{\nu}}\Lambda \indices{^{\mu '}_{\mu}} \partial_{\nu} V^\mu. 
\end{align}
In tensorial language, one says that $\partial_{\nu}V^{\mu}$ transforms as a tensor of rank (1,1), where $(n,m)$ denote $n$ indices on the top and $m$ indices on the bottom. In curved spacetime, this is no longer true because $ \Lambda \indices{_{\nu '}^{\nu}} \rightarrow \tfrac{\partial x^{\nu '}}{\partial x^{\nu}}$ is generically a spacetime dependent quantity. The partial differential operator must therefore be generalised (to a covariant derivative), so that the `differentiated' object is also a tensor. This requires the notion of parallel transport.

\subsection{Parallel transport and affine connection}
Mathematically, one has to be careful when taking a derivative of a vector because the definition of a vector along a curve defined by $\lambda$ 
\begin{align}
\frac{d V^\mu}{d\lambda} = {\rm lim}_{\delta\lambda \rightarrow 0} \frac{V^\mu(Q) - V^\mu(P)}{\delta\lambda},
\end{align}
where $Q$ and $P$ are spacetime points at $\lambda + \delta\lambda$ and $\lambda$ respectively, require comparison between two vectors in different tangent spaces. This is okay in flat spaces, but in curved spaces there is no intrinsic way to do it. What we need is a concept of parallel transport that moves a vector--or more generally a tensor--along a curve while keeping it `constant'. $V^\mu(P)$ can be parallel transported to $Q$ with the help of the `affine connection' $\Gamma$ such that
\begin{align}
\label{paralleltransport}
V^\mu(P\rightarrow Q) = V^\mu(P) - \delta\Lambda \Gamma^\mu_{\nu\sigma}(P) V^\nu(P)\frac{dx^\sigma}{d\lambda}.
\end{align}
The corresponding covariant derivative is written as
\begin{align}
\nabla_\sigma V^\mu = \frac{V^\mu(Q) - V^\mu(P\rightarrow Q)}{\delta x^\sigma} = \partial_\sigma V^\mu + \Gamma^\mu_{\nu\sigma}V^\nu,
\end{align}
with which we can define the parallel transport condition as
\begin{align}
\frac{dx^\sigma}{d\lambda}\nabla_\sigma V^\mu = 0.
\end{align}
Note that this is a generalisation of the condition in flat space $\tfrac{\partial V^\mu}{\partial x^\sigma} = 0$. Covariant derivative of a one-form, i.e.~a (0,1) tensor $\omega_\mu$ is found to be
\begin{align}
\nabla_\sigma \omega_{\nu} = \partial_\sigma \omega_\mu - \Gamma^\lambda_{\sigma\nu}\omega_\lambda.
\end{align}

The transformation properties of a connection can be found by demanding that $\nabla_\sigma V^\mu$ transforms as a tensor, i.e.,
\begin{align}
\nabla_{\sigma '} V^{\mu '} = \frac{\partial x^\sigma}{\partial x^{\sigma '}} \frac{\partial x^{\mu '}}{\partial x^{\mu}} \nabla_\sigma V^\mu.
\end{align}
This yields
\begin{align}
\Gamma^{\mu'}_{\nu'\sigma'} = \frac{\partial x^{\nu}}{\partial x^{\nu'}}\frac{\partial x^{\sigma}}{\partial x^{\sigma'}}\frac{\partial x^{\mu'}}{\partial x^{\mu}}\Gamma^{\mu}_{\nu\sigma} - \frac{\partial x^{\nu}}{\partial x^{\nu'}}\frac{\partial x^{\sigma}}{\partial x^{\sigma'}}\frac{\partial}{\partial x^\sigma}\left( \frac{\partial x^{\mu'}}{\partial x^{\nu}}\right),
\end{align}
showing that $\Gamma$ is not a tensor.

\subsection{The metric connection} 
The above consideration shows that the concept of parallel transport requires an affine connection to be defined. This is a very general property irrespective of the detailed structure of the manifold, allowing many distinct definitions of parallel transport. In general relativity, however, it is possible define a unique connection compatible with the metric $g_{\mu\nu}$ (remember, $ds^2 = g_{\mu\nu}dx^\mu dx^\nu$) as follows. First, the connection is assumed to be torsion-free, meaning that $\Gamma^\lambda_{\mu\nu} = \Gamma^\lambda_{\nu\mu}$. Second, the metric is assumed to obey the parallel transport condition: $\nabla_\sigma g_{\mu\nu} = 0$. The latter guarantees that the scalar product of two parallel-transported vectors is constant. That is, if $\tfrac{dx^\sigma}{d\lambda}\nabla_\sigma U^\mu = \tfrac{dx^\sigma}{d\lambda}\nabla_\sigma V^\nu = 0$, then $\tfrac{dx^\sigma}{d\lambda}\nabla_\sigma(g_{\mu\nu}U^\mu V^\nu) = 0$.

From the second assumption, we have the following three relations
\begin{align}
\nabla_\rho g_{\mu\nu} = \partial_\rho g_{\mu\nu} - \Gamma^{\lambda}_{\rho\mu}g_{\lambda\nu}- \Gamma^{\lambda}_{\rho\nu}g_{\mu\lambda} = 0, \nonumber \\
\nabla_\mu g_{\nu\rho} = \partial_\mu g_{\nu\rho} - \Gamma^{\lambda}_{\mu\nu}g_{\lambda\rho}- \Gamma^{\lambda}_{\mu\rho}g_{\nu\lambda} = 0, \nonumber \\
\nabla_\nu g_{\rho\mu} = \partial_\nu g_{\rho\mu} - \Gamma^{\lambda}_{\nu\rho}g_{\lambda\mu}- \Gamma^{\lambda}_{\nu\mu}g_{\rho\lambda} = 0.
\end{align}
Subtracting the second and third lines from the first and using the first assumption, we obtain
\begin{align}
\partial_\rho g_{\mu\nu} - \partial_\mu g_{\nu\rho} - \partial_\nu g_{\rho\mu} + 2\Gamma^{\lambda}_{\mu\nu}g_{\lambda\rho} = 0,
\end{align}
which after some rearranging yields
\begin{align}
\Gamma^\sigma_{\mu\nu} = \frac{1}{2}g^{\sigma\rho}\left( \partial_\mu g_{\nu\rho} + \partial_\nu g_{\rho\mu} -  \partial_\rho g_{\mu\nu}  \right).
\end{align}
This is the metric connection, called the Christoffel symbol.

\subsection{Spin connection}
So far, we have seen how to take covariant derivatives of tensors. However, this is not enough to write down the Dirac equation in a curved spacetime. We also need to know how to take the covariant derivative of a spinor. For this purpose, we will use the fact that locally it is always possible (due to the equivalence principle) to find an inertial coordinate system in which the metric becomes Minkowskian. Suppose that $y^a$ are such local coordinates at point $x^\mu = X^\mu$ (we use the convention that latin indices $a,b,...$ are used to label local inertial coordinates and greek indices $\mu,\nu,...$ for general coordinates). Then defining 
\begin{align}
e\indices{^\mu_a}(X) &= \frac{\partial x^\mu}{\partial y^a} \Bigg |_{x^\mu = X^\mu} \;\;\;\;\;  {\rm and} \nonumber \\
 e\indices{^a_\mu}(X) &= \frac{\partial y^a}{\partial x^\mu} \Bigg |_{x^\mu = X^\mu}
\end{align}
at each point in spacetime, we get the {\it vielbein} (`many-legs') $e^\mu_a(x)$ that diagonalises the metric, as well as its inverse
\begin{align}
e\indices{^\mu_a}(x)e\indices{^\nu_b}(x)g_{\mu\nu}(x) &= \eta_{ab}, \nonumber \\
 e\indices{^a_\mu}(x)e\indices{^b_\nu}(x)\eta_{ab}(x) &= g_{\mu\nu}.
\end{align}

What we want is to know what the parallel transport equation (\ref{paralleltransport}) looks like in the local inertial coordinate system. Basically, we want 
\begin{align}
\label{spintransport}
V^a(x \rightarrow x + dx) = V^a(x) - \omega\indices{^a_b_\nu}V^b(x)dx^\nu,
\end{align}
where $\omega\indices{^a_b_\nu}$ is a generalisation of the affine connection called the spin connection. Noting that
\begin{align}
V^\mu(x) &= e\indices{^\mu_a}(x)V^a(x), \nonumber \\ 
V^\mu(x\rightarrow x+dx) &= e\indices{^\mu_a}(x+dx)V^a(x\rightarrow x+dx),
\end{align}
and $e\indices{^\mu_a}(x+dx) \approx e\indices{^\mu_a}(x) + \partial_\nu (e\indices{^\mu_a}(x))dx^\nu$, we can obtain the spin connection in terms of the affine connection
\begin{align}
\label{spinconnection}
\omega\indices{^a_b_\nu} = e\indices{^a_\mu} \partial_\nu (e\indices{^\mu_b}) + e\indices{^a_\mu}e\indices{^\sigma_b}\Gamma^\mu_{\sigma\nu}.
\end{align}
This allows one to take the covariant derivative of a tensor with mixed indices. For example
\begin{align}
\nabla_\nu e\indices{^\mu_a} = \partial_\nu e\indices{^\mu_a} + \Gamma^\mu_{\sigma\nu}e\indices{^\sigma_a} - \omega \indices{^b_a_\nu} e\indices{^\mu_b} = 0,
\end{align}
where the last equality follows from (\ref{spinconnection}). The spin connection is antisymmetric in the first two indices, i.e., $\omega_{ab\nu} = -\omega_{ba\nu}$, so that the magnitude of the Lorentz vector remains constant upon parallel transport.

Using the spin connection we can derive the covariant derivative operator for spinors. The latter is defined through the parallel transport equation
\begin{align}
\psi(x\rightarrow x + dx) = \psi(x) - \Omega_\nu(x)\psi(x)dx^\nu,
\end{align}
where $\Omega_\nu$ is an n-by-n matrix for each index $\nu$ with n=2 for 2 and 3 spacetime dimensions and 4 for 4-dimensional spacetime. To determine $\Omega_\nu$, we use the fact that $S(x) = \bar{\psi}(x)\psi(x)$ and $V^a(x) = \bar{\psi}(x)\gamma^a\psi(x)$ transform as a scalar and a vector, respectively. 
Firstly,
\begin{align}
S(x\rightarrow & x + dx) - S(x)  \nonumber \\  &=  \bar{\psi}(x)\left[ \gamma^0\Omega^\dagger_\nu(x)\gamma^0 + \Omega_\nu(x)\right]\psi(x) dx^\nu,
\end{align}
yielding
\begin{align}
\label{cond1}
\gamma^0\Omega^\dagger_\nu(x)\gamma^0 =- \Omega_\nu(x).
\end{align}
Secondly,
\begin{align}
V^a(x\rightarrow & x + dx) - V^a(x) \nonumber \\ &= - \bar{\psi}(x)\left[ \gamma^a\Omega_\nu +\gamma^0\Omega^\dagger_\nu\gamma^0\gamma^a\right]\psi(x) dx^\nu, \nonumber \\
&= - \bar{\psi}(x)\left[ \gamma^a\Omega_\nu - \Omega_\nu\gamma^a\right]\psi(x) dx^\nu, \nonumber \\
&= - \omega\indices{^a_b_\nu}\gamma^bdx^\nu.
\end{align}
The second equality results from (\ref{cond1}), while the third equality results from the definition of the spin connection. From the second and third lines we conclude that
\begin{align}
\label{cond2}
[\gamma^a,\Omega_\nu] = \omega\indices{^a_b_\nu}\gamma^b.
\end{align}
Noting that
\begin{align}
[\gamma^a,\sigma^{bc}] = 2i(\eta^{ab}\gamma^c - \eta^{ac}\gamma^b),
\end{align}
where $\sigma^{bc} = i[\gamma^b,\gamma^c]/2$, one can verify by direct substitution that
\begin{align}
\Omega_\nu(x) = -\frac{i}{4}\omega_{ab\nu}(x)\sigma^{ab}
\end{align}
satisfies (\ref{cond2}), while (\ref{cond1}) can be verified using the relationship $\gamma^0\gamma^{a\dagger}\gamma^0 = -\gamma^a$.

Using the covariant derivative $\nabla_\nu = \partial_\nu + \Omega_\nu$,  the flat spacetime Dirac equation (c=1)
\begin{align}
\left[i\gamma^a\partial_a\psi -m \right] \psi(x) = 0,
\end{align}
generalises to the curved spacetime Dirac equation
\begin{align}
\label{curveddiraceqn}
\left[ i\gamma^\mu\nabla_\mu -m\right]\psi(x) = 0,
\end{align}
where the vielbein was used to transform the local $\gamma$ matrices: $\gamma^\mu = e\indices{^\mu_a}(x)\gamma^a$. Note that $\{ \gamma^\mu(x),\gamma^\nu(x) \} = 2g^{\mu\nu}(x)$. From here on, we will use the notation $\tilde{\gamma}^a \equiv \gamma^a$ to avoid confusion when numerical indices are substituted. 

\section{Dirac equation in 2 dimensional curved spacetime}
A special feature of the 2 dimensional spacetime is that the metric can always be reduced to the conformally flat form
\begin{equation}
ds^2=\Omega^2(dt^2-dx^2)
\end{equation}
for some function $\Omega(x,t)$.
To derive the Dirac equation in this metric we first need the Christoffel symbols which are readily calculated to be:
$\Gamma^0_{00}=\Gamma^0_{11}=\Gamma^1_{10}=\Gamma^1_{01}=\dot{\Omega}/\Omega$ and 
$\Gamma^0_{01}=\Gamma^0_{10}=\Gamma^1_{00}=\Gamma^1_{11} =\Omega'/\Omega$.
The dot denotes a derivative with respect to time and the prime a spatial derivative. Using these, the vielbein can be readily calculated:  $e\indices{^0_0}= e\indices{^1_1}= 1/\Omega$. 
These lead to non-vanishing spin connections
$\omega\indices{^0_1_0}=\omega\indices{^1_0_0}= \Omega'/\Omega$ and $\omega\indices{^0_1_1}=\omega\indices{^1_0_1}= \dot{\Omega}/\Omega$,
which in turn lead to $\Omega_0= \Omega'/(4\Omega) [\tgamma^0,\tgamma^1]$ and $\Omega_1= \dot{\Omega}/(4\Omega) [\tgamma^0,\tgamma^1]$. 

That this spacetime is curved can be verified by calculating the Ricci curvature $R = g^{\mu\nu}R_{\mu\nu} = 2\left( (\dot{\Omega}/\Omega)^2 - \ddot{\Omega}/\Omega\right)/\Omega^2$, where $R_{\mu\nu} = \partial_\lambda \Gamma^{\lambda}_{\mu\nu} - \partial_\nu \Gamma^{\lambda}_{\mu\lambda} + \Gamma^{\lambda}_{\mu\nu}\Gamma^{\sigma}_{\lambda\sigma}- \Gamma^{\lambda}_{\mu\sigma}\Gamma^{\sigma}_{\nu\lambda}$ is the Ricci tensor. 

We can thus write down the Dirac equation in a general 1+1 dimensional spacetime. Inserting the above results into Eq.~(\ref{curveddiraceqn}), multiplying with $\Omega \tgamma^0$ from the left, and rearranging, we get
\begin{equation}
\begin{split}
i(\partial_t+\tgamma^0\tgamma^1\frac{[\tgamma^0,\tgamma^1]}{4}\frac{\dot{\Omega}}{\Omega})\psi &= i[\tgamma^0\tgamma^1\partial_x + \frac{[\tgamma^0,\tgamma^1]}{4} \frac{\Omega'}{\Omega} ]\psi \\
&\quad+\tgamma^0\Omega m\psi.
\end{split}
\end{equation}
Choosing $\tgamma^0 = \sigma_z$ and $\tgamma^1=  i\sigma_y$, the Dirac equation becomes
\begin{equation}
i\left(\partial_t+\frac{\dot{\Omega}}{2\Omega}\right)\psi= \left[-i\sigma_x\left(\partial_x + \frac{\Omega'}{2\Omega} \right)+\sigma_z\Omega m\right]\psi(x).
\end{equation}
Which, upon defining $\sqrt{\Omega} \xi=  \psi$, can be recast into our final form:
\begin{equation}
\label{curvedDirac}
i\partial_t(\Omega\xi)= -i\sigma_x\partial_x(\Omega\xi)+\sigma_z \Omega m (\Omega\xi).
\end{equation}
$\Omega \xi$ thus solves the regular Dirac equation with an effective mass of $\Omega m$ which, as we show below, means that it can be simulated in binary waveguide arrays. Note that in flat spacetime $\Omega = 1$ and the equation reduces to
\begin{align}
\label{flateqn}
i\partial_t \psi= -i\sigma_x\partial_x\psi+\sigma_z  m \psi.
\end{align}

\subsection{Specific examples}
Here we provide examples of specific spacetimes and corresponding Dirac equations.

\subsubsection{Static spacetime}
Static spacetimes are described by a metric that has a time-like variable $x_0$ such that $g_{0i} = 0$ and $\partial_0 g_{\mu \nu} = 0$.
In 1+1 dimensions the line element for these spacetimes may be written as
\begin{equation}
ds^2=e^{2\Phi}dt^2-e^{2\Psi}dx^2,
\end{equation}
where $\Phi$ and $\Psi$ are independent of $t$.
Non-vanishing Christoffel symbols for this metric are $\Gamma^0_{10}=\Gamma^0_{01}= \Phi'$, $\Gamma^1_{00}= \Phi'e^{2(\Phi-\Psi)}$, and $\Gamma^1_{11}= \Psi'$, whereas 
the vielbein is easily found to be $e\indices{^0_0}= e^{-\Phi}$, $e\indices{^1_1}= e^{-\Psi}$.
Then the non-vanishing spin connections are
$\omega\indices{^1_0_0}=\omega\indices{^0_1_0}= \Phi'e^{\Phi-\Psi}$,
leaving us with the non-vanishing element of $\Omega_\nu$:
$\Omega_0 = \frac14\Phi'e^{\Phi-\Psi}[\tgamma^0,\tgamma^1]$.
Inserting the results and using the same gamma-matrices as above, we can write the Dirac equation as
\begin{equation}
i\partial_t \psi =-ie^{\Phi-\Psi}\sigma_x\left(\partial_x+\frac{\Phi'}{2} \right)\psi + e^{\Phi}\sigma_z m\psi.
\end{equation}

\subsubsection{FRW Metric}
The FRW metric in 1+1 D reads
\begin{equation}
ds^2=dt^2-a^2(t) dx^2.
\end{equation}
The non-vanishing spin connections, Christoffel symbols, and $\Omega_{\nu}$ are
$e\indices{^0_0}= 1$, $e\indices{^1_1}= 1/a$;
$\Gamma^0_{11}= a\dot{a}$; $\Gamma^1_{10}=\Gamma^1_{01}= \dot{a}/{a}$;
$\omega\indices{^0_1_1}=\omega\indices{^1_0_1}= \dot{a}$; $\Omega_1= \frac{\dot{a}}{4}[\tgamma^0,\tgamma^1]$.
The Dirac equation reads
\begin{equation}
i(\tgamma^0 \partial_0 \psi +  \tgamma^1\frac1a \partial_x \psi+\tgamma^1\frac{\dot{a}}{4a}[\tgamma^0,\tgamma^1] \psi)-m\psi = 0.
\end{equation}
After plugging in $\tgamma^a$, we obtain
\begin{equation}
i\partial_t \psi = -i\frac{\dot{a}}{2a} \psi -i\frac{\sigma_x}{a} \partial_x \psi+\sigma_z m\psi.
\end{equation}
Note that the FRW metric can be converted to the conformally flat form by setting $\eta(t) = \int^t \frac{dt}{a(t)}$.

\subsubsection{Rindler spacetime}
Rindler metric describes the dynamics of a uniformly accelerating observer in flat spacetime. Even though the spacetime is flat, Unruh showed that spontaneous particle creation occurs in the frame of the accelerating observer \cite{Unruh1976} analogously to the famous Hawking radiation \cite{Hawking1975}. The vielbein formalism described above proves useful for deriving the wave equation in this metric. By introducing the coordinate $t = u\sinh v$ and $x = u \cosh  v$, the Minkowski metric is converted to \cite{Soffel1980}
\begin{align}
ds^2 = u^2dv^2 - du^2,
\end{align}
when $u$ is constrained to be positive. Because this is a static spacetime, we can use the formula derived above to obtain the Dirac equation in the Rindler spacetime:
\begin{align}
i\partial_v\psi = -iu\sigma_x\left( \partial_u +\frac{1}{2u} \right)\psi + mu\sigma_z\psi.
\end{align}

The metric can also be put in a conformally flat form by introducing new coordinates
\begin{align}
t = \frac{1}{a}e^{a\xi}\sinh (a\eta), \;\;\; x = \frac{1}{a}e^{a\xi}\cosh (a\eta),
\end{align}
in the region $x>|t|$. In these coordinates, the metric takes the form
\begin{align}
ds^2 = e^{2a\xi}\left( d\eta^2 - d\xi^2\right),
\end{align}
and the Dirac equation reads
\begin{align}
i\partial_\eta\psi = \left[ -i\sigma_x\left( \partial_\xi + \frac{1}{2}\right) + \sigma_z e^{a\xi}m \right]\psi.
\end{align}

\subsection{Particle creation in curved spacetime}
\label{sect3b}
It is well established that there is particle creation in curved spacetime in general \cite{Birrell, Fulling, Mukhanov, Parker}. The background metric acts in a similar manner to an external field such as, for example, the electromagnetic field, and something akin to the Schwinger effect (creation of electron-positron pair in strong electric fields) occur. Consider as an example an expanding-universe scenario where the expansion is asymptotically turned on. The field is initially in the vacuum state and as expansion is gradually turned on, electrons and positrons pop out in pairs. The essence of understanding this phenomena is that a vacuum state is not unique. It is defined as an eigen-state of the field operator $\psi$ with eigenvalue 0. To define the latter more precisely, one has to expand the field operator in terms of mode functions and assign creation and annihilation operators that create and annihilate these modes. The vacuum is then the state with eigenvalue 0 for all the mode-annihilation operators. Now, these mode functions depend on the background spacetime, which means that the vacuum state of one spacetime need not be the vacuum state of another. Below, we provide a more detailed and pedagogical explanation of this effect for scalar fields in FRW spacetime, before we come back to Dirac fermions.

\subsubsection{Scalar field in FRW spacetime}
A scalar field obeying the Klein-Gordon equation is one of the simplest quantum fields to deal with and a spatially-flat, time-dependent metric is one of the simplest examples of curved spacetime metrics. As an example of such a metric, we choose the FRW metric and explain creation of scalar fields in it. We follow closely the exposition by Mukhanov and Winitzki \cite{Mukhanov}, but work in 2 dimensional spacetime instead of the usual 4 dimensional one. Remember that a  real scalar field in Minkowski (or flat) spacetime obeys the Klein-Gordon equation
\begin{align}
\partial_\mu\partial^\mu \phi + m^2\phi = 0.
\end{align}
In curved spacetime this becomes (in the minimal coupling scheme)
\begin{align}
g^{\mu\nu}\nabla_\mu\nabla_\nu \phi + m^2\phi = 0.
\end{align}
To work out the explicit form of this equation, it is helpful to convert it to the following form
\begin{align}
g^{\mu\nu}\partial_\mu\partial_\nu\phi + \frac{1}{\sqrt{-g}}\left( \partial_\nu\phi\right) \partial_\nu \left( g^{\mu\nu}\sqrt{-g}\right) + m^2\phi = 0,
\end{align}
where $g$ is the determinant of the metric. Note that $g$ depends on the spacetime dimension: for the case of conformal spacetime described by $ds^2 = \Omega^2(dt^2 - dx^2 \cdots )$, $g = \Omega^2d$, where $d$ is the dimension of the spacetime. Working with the conformal version of the FRW metric, the wave equation evaluates to
\begin{align}
\ddot{\phi} - \phi '' + \Omega(\eta)^2m^2\phi = 0
\end{align}
in 1+1D and
\begin{align}
\ddot{\phi} + 2\frac{\dot{\Omega}}{\Omega}\dot{\phi}- \nabla\phi + \Omega(\eta)^2m^2\phi = 0
\end{align}
in 3+1D. As before, the dot denotes derivative with respect to the conformal time $t$, whereas the prime denotes a spatial derivative. 

Going to the momentum space by defining
\begin{align}
\phi(x,t) = \frac{1}{\sqrt{2\pi}}\int dk \phi_k(t) e^{ikx},
\end{align}
the wave equation reduces to
\begin{align}
\ddot{\phi}_k - k^2\phi_k + \Omega^2(t) m^2\phi_k \equiv \ddot{\phi}_k + \omega_k^2(t)\phi_k = 0.
\end{align}
The general solution of this time-dependent oscillator equation can be written in terms of complex mode functions $v_k(t)$ with the normalisation condition $-iW[v_k,v_k^*]/2 \equiv -i(\dot{v}_k v_k^* - v_k\dot{v}_k^*)./2= 1$, where $W$ is called the Wronskian. The field operator $\phi_k$ can be expanded as
\begin{align}
\phi_k(t) = \frac{1}{2}\left[ a_k v_k^*(t) + a^\dagger_k v_k(t) \right],
\end{align}
where $a_k$ and $a^\dagger_k$ are the bosonic annihilation and creation operators for mode $k$. It is interesting to note that the (bosonic) commutation relation is crucial for preserving the normalisation defined by the Wronskian. The anticommutation relation (for fermions) is not consistent with the latter. 

Once the mode expansion is defined, the vacuum state $|0\rangle$ is defined by the condition $a_k|0\rangle = 0$ for all $k$. Now consider two different mode functions $u_k(t)$ and $v_k(t)$. Since $u_k$ and $u_k^*$ form a basis, we can expand $v_k$ in terms of them, $v_k^*(t) = \alpha_k u_k^*(t) + \beta_k u_k(t)$, with $t$-independent coefficients $\alpha_k$ and $\beta_k$ that obey the normalisation condition $|\alpha_k|^2 - |\beta_k|^2 = 1$. If $b_k$ and $b_k^\dagger$ are the annihilation and creation operators corresponding to the mode functions $u_k$, the following relations hold
\begin{align}
b_k = \alpha_ka_k + \beta_k^* a_{-k}^\dagger, \;\;\;\;\; b_k^\dagger = \alpha_k^*a_k^\dagger + \beta_k a_{-k}.
\end{align}
This is called the Bogolyubov transformation in the literature. What is interesting is that the vacuum state for $a$ is not the vacuum state for $b$. Indeed a simple calculation reveals that the expectation value of the number of $b$-particles in $a$'s vacuum state is generically non-zero:
\begin{align}
\langle 0_a | b_k^\dagger b_k |0_a\rangle &= \langle 0_a | (\alpha_k^*a_k^\dagger + \beta_k a_{-k}) (\alpha_k a_k + \beta_k^*a_{-k}^\dagger) |0_a\rangle \nonumber \\
&= |\beta_k|^2 \langle 0_a | a_{-k}a_{-k}^\dagger |0_a\rangle = |\beta_k|^2.
\end{align}
To repeat, if $\beta_k$ is non-zero, the vacuum state defined with respect to the $a_k$ modes contains a finite density of $b_k$ particles.

It is generally impossible to find a unique vacuum state in curved spacetime unlike in flat (Minkowski) spacetime. To discuss particle creation, it is therefore a good idea to work with an asymptotically flat spacetime that has the Minkowski metric both in far-enough past and far-enough future. As a simple example, we here consider a specially chosen FRW spacetime with $\Omega^2(t) = 1$ if $t<0$ or $t>t_0$ and $\Omega^2(t)=-1$ if $0<t<t_0$. The `in' and `out' vacuum modes at $t<0$ and $t>t_0$ are described by the standard Minkowski mode functions
\begin{align}
v_k^{(in)}(t) = \frac{1}{\sqrt{\omega_k}}e^{i\omega_kt},\;\;\;
v_k^{(out)}(t) = \frac{1}{\sqrt{\omega_k}}e^{i\omega_k(t-t_0)},
\end{align}
with $\omega_k = \sqrt{k^2 + m^2}$. To find the Bogolyubov coefficients, we need to find the relationship between the input and output modes. Solving the mode equations in all regions, one obtains the following relation for $t>t_0$
\begin{align}
v_k^{(in)}(t) = \frac{1}{\sqrt{\omega_k}}\left[ \alpha_k^* e^{i\omega_k(t-t_0)}+\beta_k^* e^{-i\omega_k(t-t_0)}\right],
\end{align}
where
\begin{align}
\alpha_k &= \tfrac{e^{-i\Omega_k t_0}}{4}\left( \sqrt{\tfrac{\omega_k}{\Omega_k}}+\sqrt{\tfrac{\Omega_k}{\omega_k}}\right)^2 -\tfrac{e^{i\Omega_k t_0}}{4}\left( \sqrt{\tfrac{\omega_k}{\Omega_k}}-\sqrt{\tfrac{\Omega_k}{\omega_k}}\right)^2, \nonumber \\
\beta_k &= \frac{1}{2}\left(\tfrac{\Omega_k}{\omega_k}-\tfrac{\omega_k}{\Omega_k}\right)\sin(\Omega_kt_0), 
\end{align}
with $\Omega = \sqrt{k^2-m^2}$. This tells us that the field mode that has evolved from $|0_{in}\rangle$ is different from the vacuum state at $t>t_0$ and has a finite particle number density
$n_k = |\beta_k|^2 = \tfrac{m^4}{|k^4-m^4|}|\sin\left( t_0(k^2-m^2)\right)|^2$

\subsubsection{Dirac field in curved spacetime, a single particle analog of particle creation}
The above procedure applies in exactly the same way for Dirac fields, except that one has to use the anticommutation relation instead of the commutation relation when quantizing the field operators. One other difference is that particles are produced in pairs so that the total charge is conserved. Because the mode expansion is a little more involved, and we are actually interested in the single-particle manifestation of pair creation, we will not go into the details here. Interested readers are referred to the original article by Parker \cite{Parker1971}. Instead let us discuss the single-particle manifestation of pair creation. In order to see how this is possible, consider the pair creation process due to an electric field (Schwinger effect) in the Dirac sea picture. Initially, the negative-energy sea is fully occupied. Then upon applying a strong electric field, a negative energy electron is kicked (or tunnels) to occupy a positive energy state, leaving a hole (positron) behind. Therefore, in terms of single-particle dynamics, what we will see is a conversion of a negative energy wave packet to a positive energy wave packet. In the next section we verify this claim, by studying the dynamics of a wave packet in curved spacetime.

\section{Dynamics of spinor wave packet}
\label{dynamics}
In this section, we provide a basic background on the wave packet dynamics, both in flat and curved spacetime. In particular, we show a single-particle analog of particle creation in an FRW spacetime. The results from this section will be used in the next section, where we discuss the implementation of wave packet dynamics in binary waveguide arrays.

\subsection{Flat spacetime}
The Dirac equation in flat 2 dimensional space time was written in Eq.~(\ref{flateqn}), which we repeat here for the reader's convenience
\begin{align}
i\partial_t \psi= -i\sigma_x\partial_x\psi+\sigma_z  m \psi.
\end{align}
Using the plane wave ansatz, the eigen-solutions can be easily found. The two solutions with positive and negative energies can be written as 
\begin{align}
\label{eestates}
\psi_+^k(x,t) &= \frac{1}{\sqrt{2 \pi}}\frac{\exp [ -iE_kt + ikx]}{\sqrt{2E_k(E_k + m)}}\begin{pmatrix} E_k+m \\ k \end{pmatrix}, \nonumber \\
\psi_-^k(x,t) &= \frac{1}{\sqrt{2 \pi}}\frac{\exp [ iE_kt + ikx]}{\sqrt{2E_k(E_k + m)}}\begin{pmatrix} -k \\ E_k + m \end{pmatrix}.
\end{align}
The time evolution of an initial spinor wave packet, $(\phi_1(k),\phi_2(k))$, is given by
\begin{align}
\psi(x,t) = \int dk \tfrac{\alpha (E_k+m)\phi_1(k) + \beta k\phi_2(k) }{2 \sqrt{2\pi} E_k(E_k + m)}\begin{pmatrix}  E_k +m \\ k \end{pmatrix} e^{i(kx-E_kt)} \nonumber \\
+\int dk \tfrac{-\alpha k \phi_1(k) + \beta (E_k+m)\phi_2(k) }{2 \sqrt{2\pi} E_k(E_k + m)}\begin{pmatrix} -k \\ E_k +m \end{pmatrix} e^{i(kx+E_kt)}.
\end{align}

\begin{figure}[ht]
\begin{center}
\includegraphics[width=0.4\columnwidth]{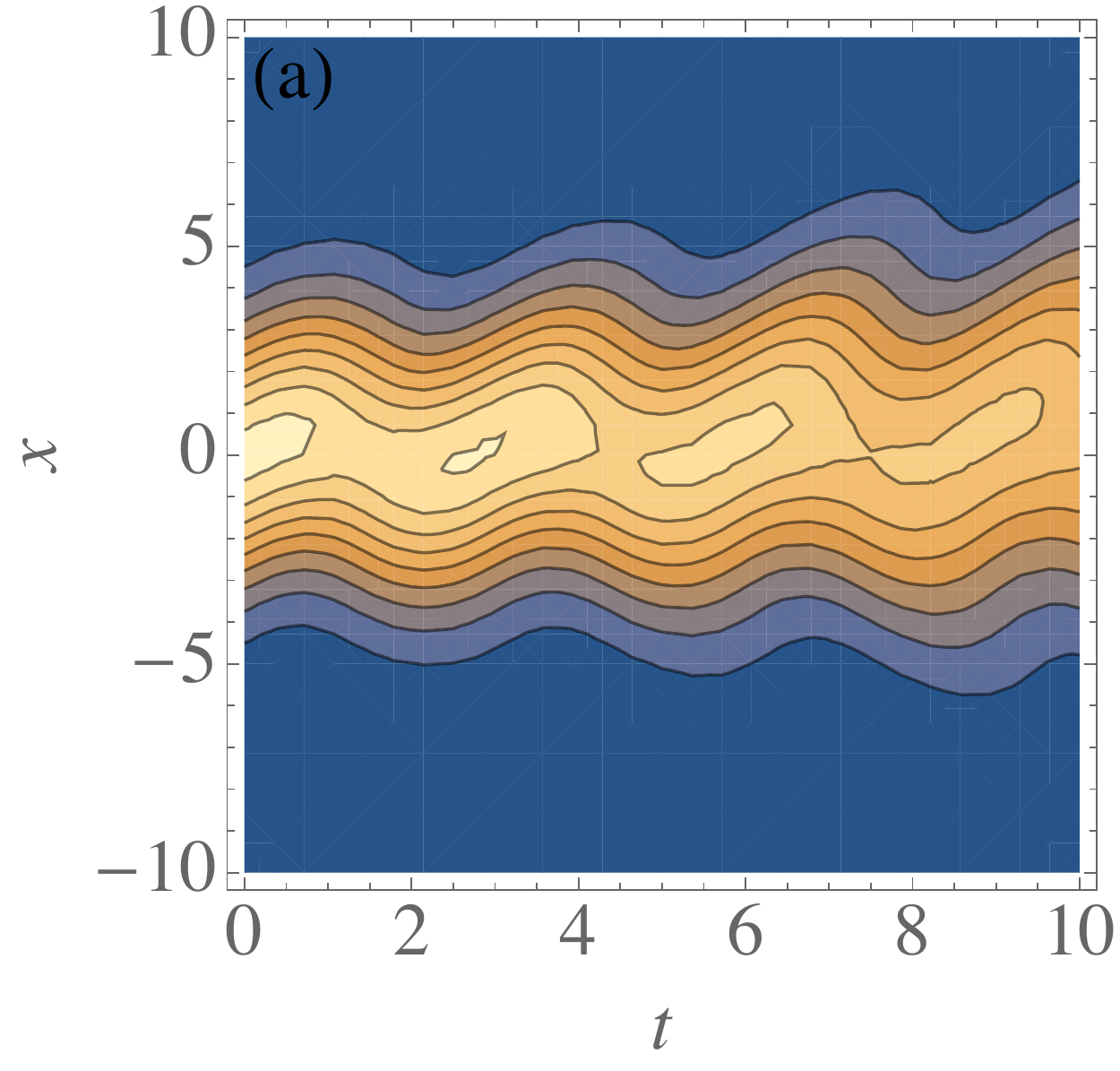}
\includegraphics[width=0.4\columnwidth]{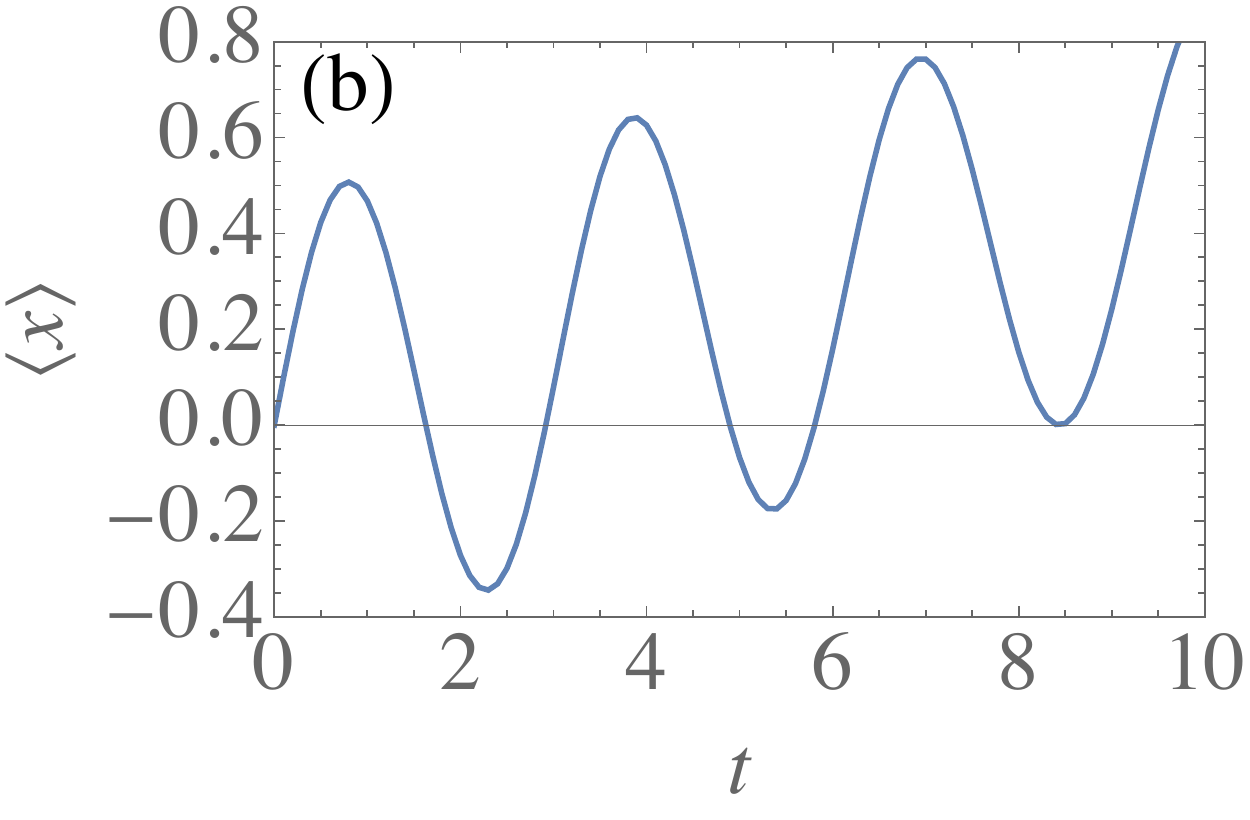}
\caption{Wave packet evolution of an initial Gaussian state, showing zitterbewegung, the trembling oscillation of the centre of mass.}
\label{flat1}
\end{center}
\end{figure}

\begin{figure}[ht]
\begin{center}
\includegraphics[width=0.4\columnwidth]{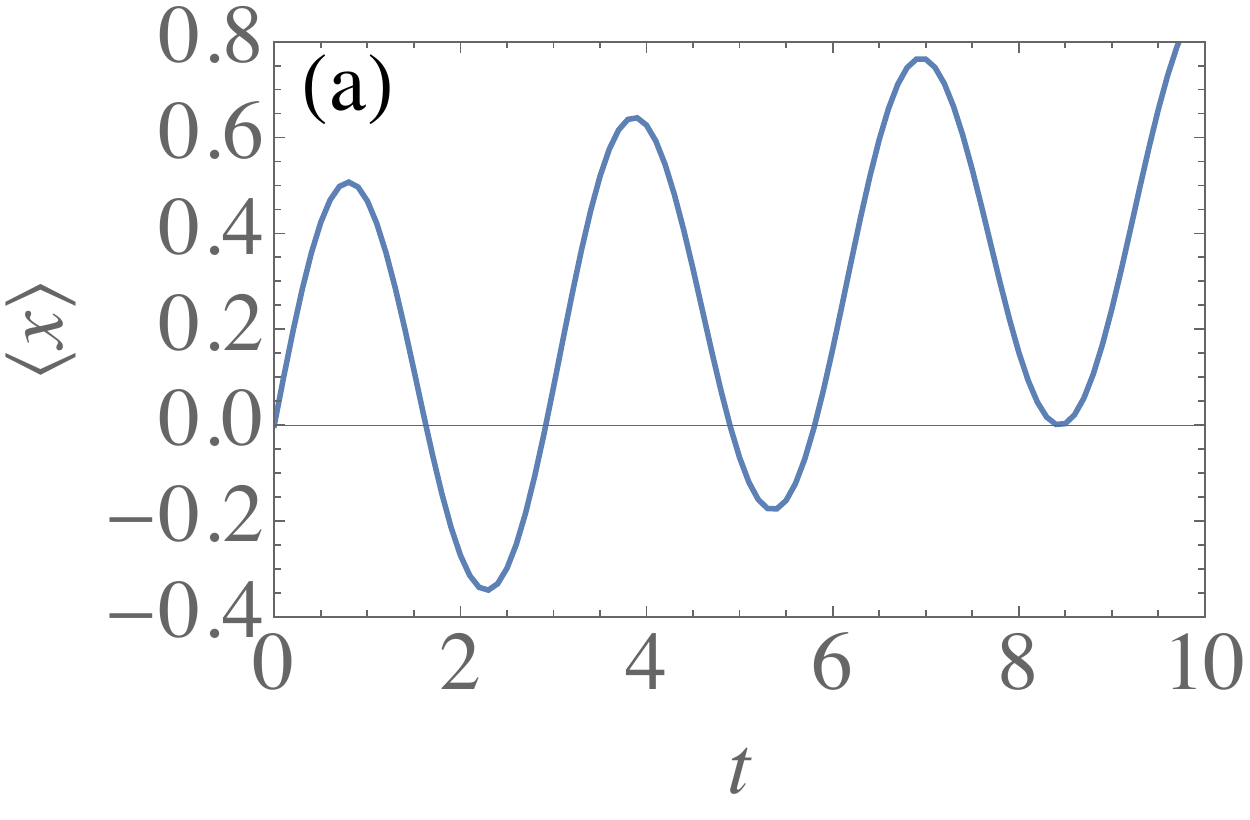}
\includegraphics[width=0.4\columnwidth]{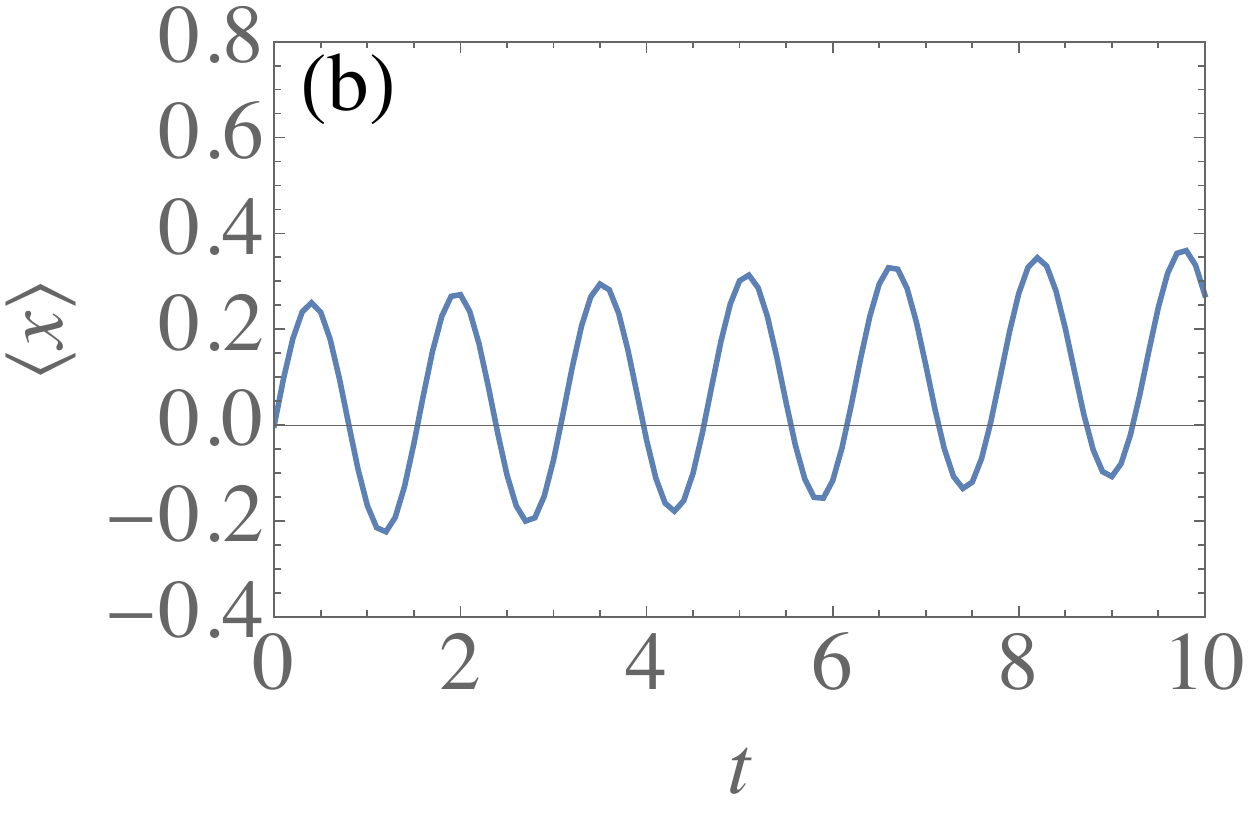}

\includegraphics[width=0.4\columnwidth]{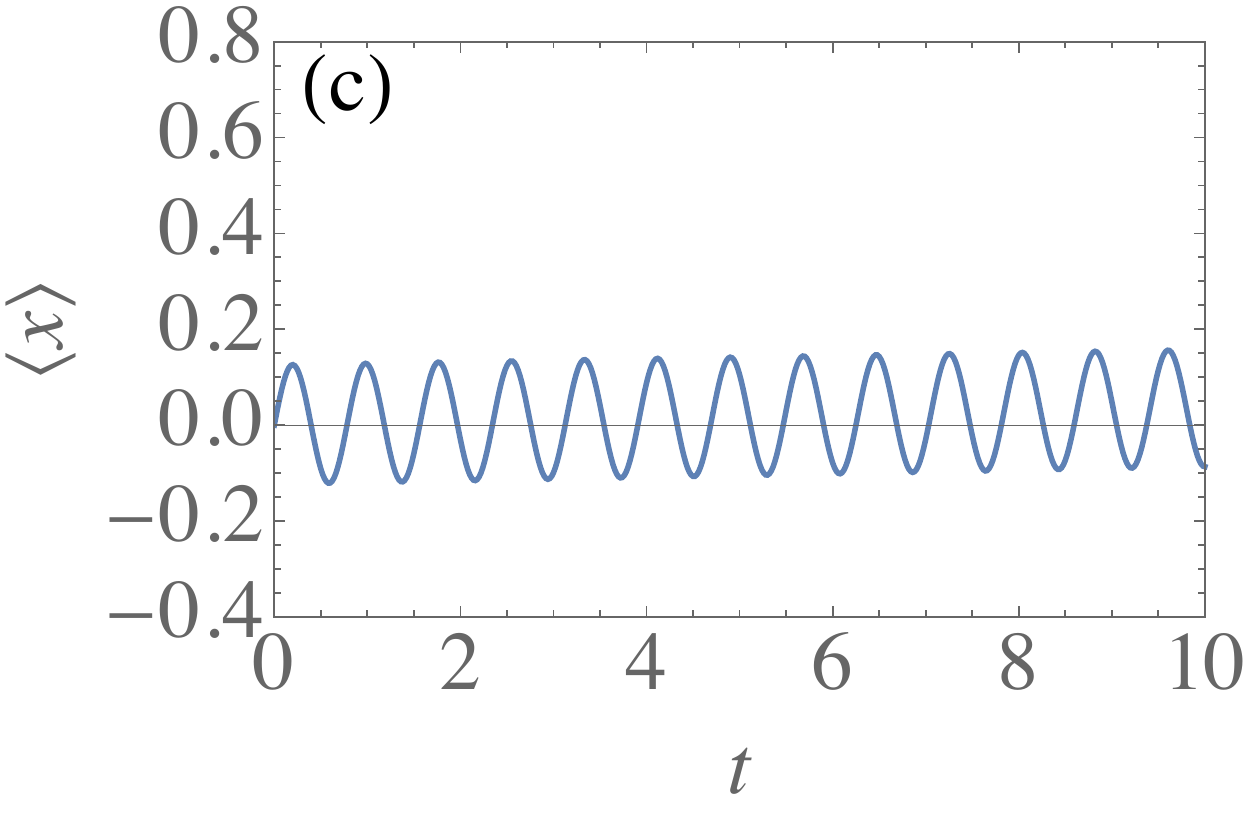}
\includegraphics[width=0.4\columnwidth]{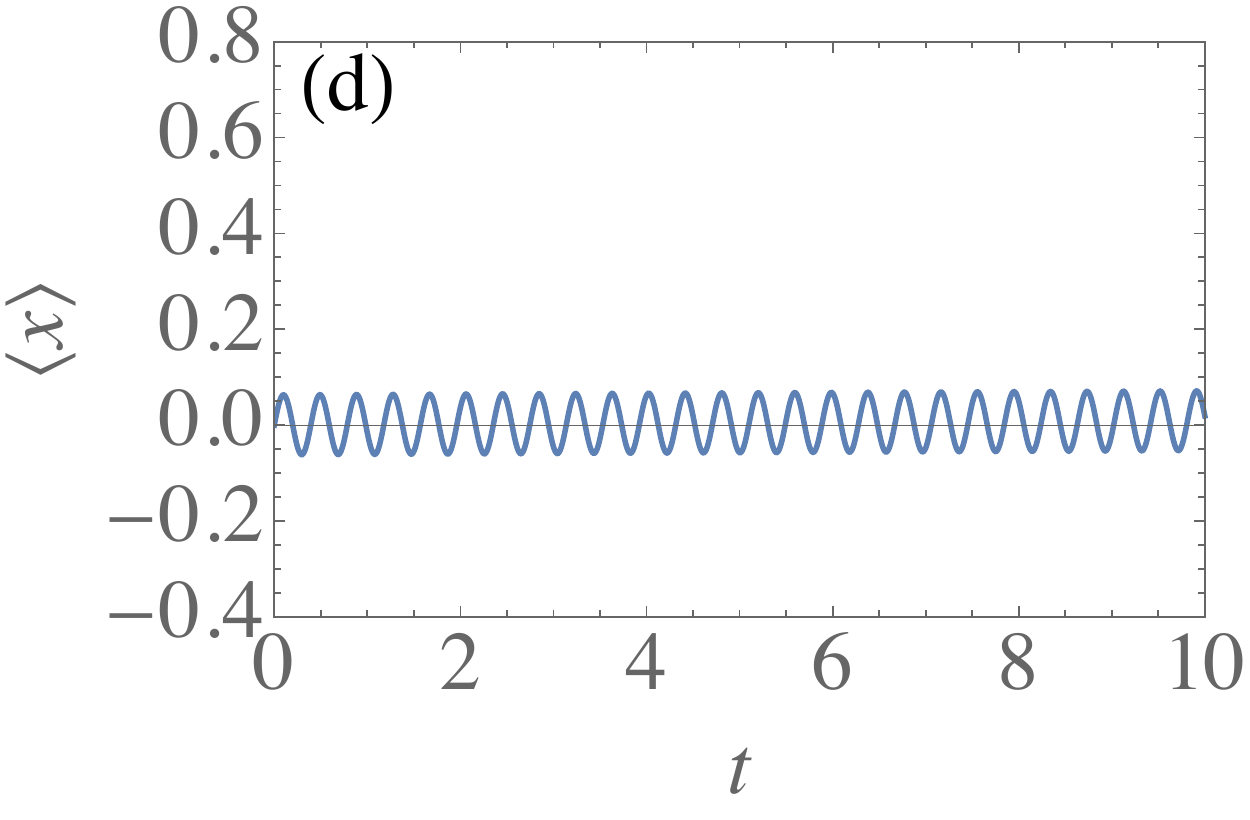}
\caption{The mass dependence of zitterbewegung for Gaussian initial states $\propto \exp[-x^2/2\sigma^2](1,1)^T$ with the width $\sigma$ = 3 and mass $m$= 1 (a), 2 (b), 4 (c), and 8 (d). }
\label{ZB}
\end{center}
\end{figure}

Let us first look at the dynamics of a Gaussian wave packet $\psi_0(x) \propto \exp[-x^2/18](1,1)^T$. Figure \ref{flat1}(a) shows the time evolution of the wave packet for $m=1$. Notice the zig-zag motion of the centre of mass as exemplified in Fig.~\ref{flat1}(b). This phenomenon is called zitterbewegung and occurs because of the interference between the positive- and negative-energy components of the spinor.  Properties of zitterbewegung can be seen from the position operator in the Heisenberg picture. Noting that $v(t) = -i[x,H] = \sigma_x(t)$ and 
\begin{align}
\frac{\partial\sigma_x}{\partial t} = -i\left[ \sigma_x,H\right] = -2i\left[p-H\sigma_x \right],
\end{align}
we obtain
\begin{align}
\sigma_x(t) = H^{-1}p + e^{2iHt}\left[ \sigma_x(0) - H^{-1}p\right].
\end{align}
$x(t) = \int v(t)$ is then given by
\begin{align}
x(t) = x(0) + \frac{p}{H}t - \frac{i}{2H}\left( e^{2iHt} - 1\right) \left[ \sigma_x(0) - \frac{p}{H}\right].
\end{align}
From the last term of this expression we notice several things. 1) the frequency of the oscillation is $2E$, 2) the amplitude is proportional to 1/(2E), and 3) zitterbewegung is non-zero only if there is a superposition between positive and negative energy states with equal momentum. The last observation follows because $\sigma_x - p/H$ anti-commutes with the Hamiltonian, which means that the matrix element of it is non-zero only between the eigenstates with equal momentum and opposite energies.  Note that for a small initial momentum, the amplitude and frequency of ZB is thus 2m and 1/(2m) respectively. This is demonstrated in Fig.~\ref{ZB}.

\begin{figure}[ht]
\begin{center}
\includegraphics[width=0.4\columnwidth]{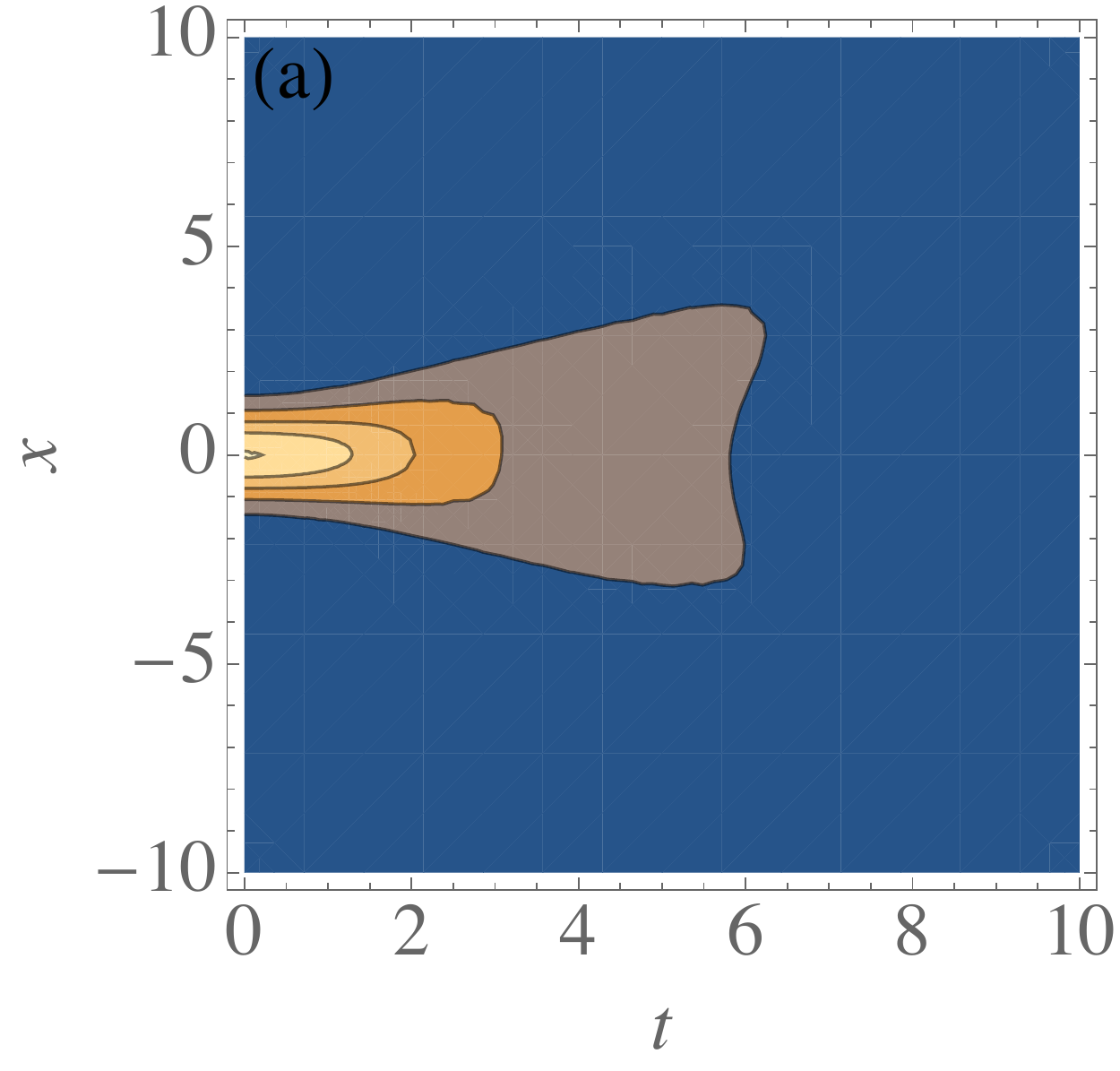}
\includegraphics[width=0.4\columnwidth]{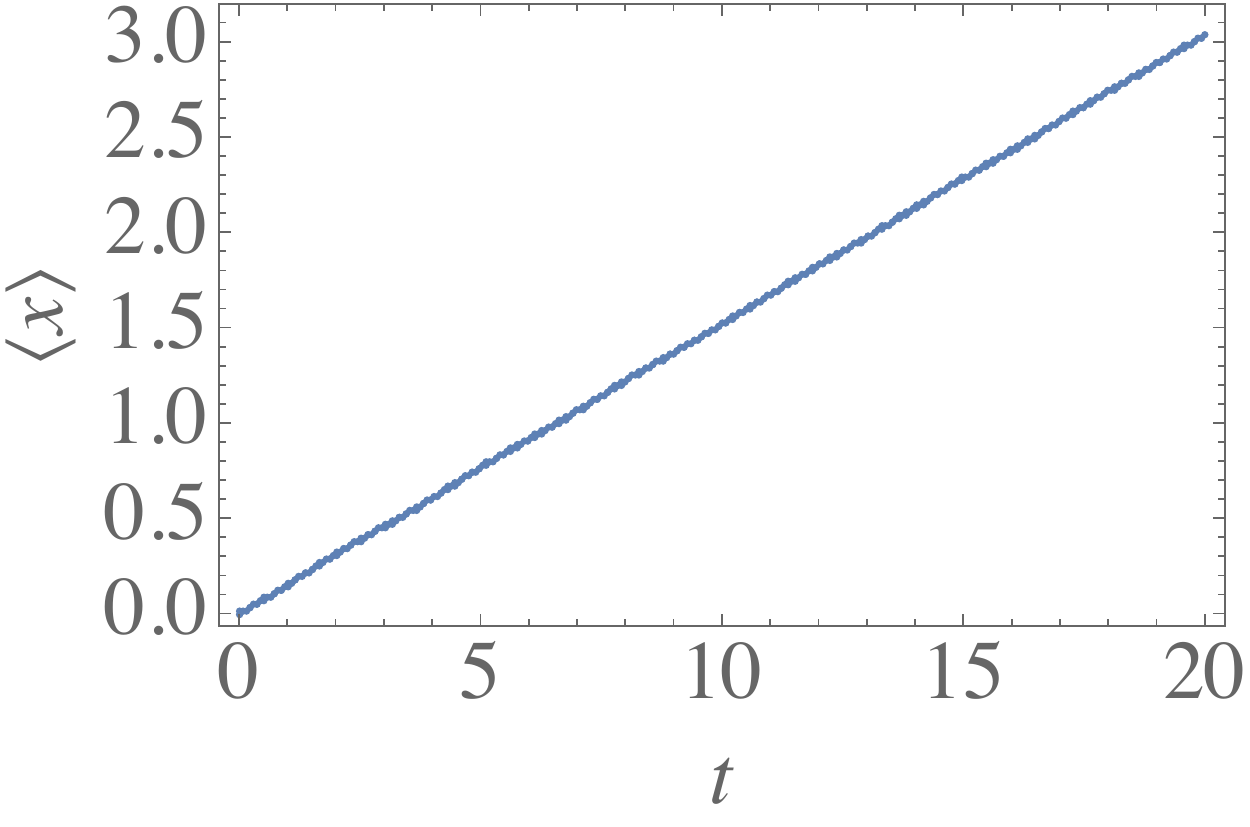}
\caption{Wave packet evolution of a positive-energy state. No zitterbewegung is observed.}
\label{flat2}
\end{center}
\end{figure}

Absence of zitterbewegung in a wave packet composed of positive-energy spinors only is shown in Fig.~\ref{flat2}. The positive energy spinor is constructed by superposing the positive-energy spinor in momentum space with a Gaussian weight $\exp[-k^2/2]$. 

\subsection{Curved spacetime}
\label{sect:curvedDirac}
We have seen that in curved spacetime, the Dirac equation is given by Eq.~(\ref{curvedDirac}):
\begin{equation*}
i\partial_t(\Omega\xi)= -i\sigma_x\partial_x(\Omega\xi)+\sigma_z \Omega m (\Omega\xi),
\end{equation*}
which we write as
\begin{equation}
i\partial_t\psi= -i\sigma_x\partial_x\psi+\sigma_z m_{eff}(t)\psi,
\end{equation}
Here we concentrate on the FRW spacetime with a time-dependent conformal factor $a(t) \equiv \Omega(t)$. Neglecting the overall factor $\Omega(t)$, this equation is simply the Dirac equation with a time-dependent mass term. Analogously to a real scalar field in an FRW spacetime, Dirac fermions are produced (in pairs) in generic cases, which we will show by demonstrating conversion of a negative energy wavepacket to a positive energy wave packet. Quantitatively, the conversion can be proven by calculating the norm of the positive-energy-projected spinor wave packet, but we can do better: We can make use of the fact that ZB only occurs when positive-energy spinors are superposed with negative-energy spinors. So our aim is to show how ZB is induced in an asymptotically flat spacetime with an initial negative-energy wave packet. 

\begin{figure}[ht]
\begin{center}
\includegraphics[width=0.45\columnwidth]{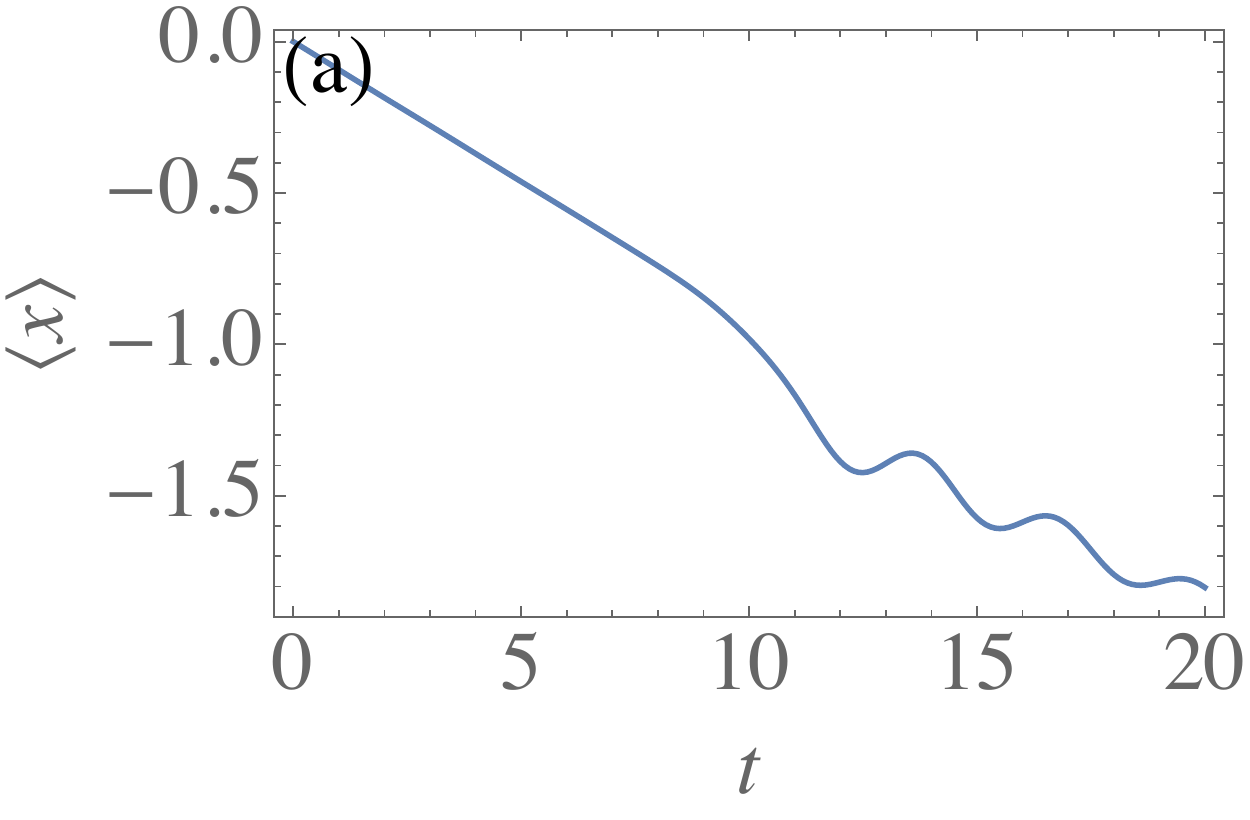}
\includegraphics[width=0.45\columnwidth]{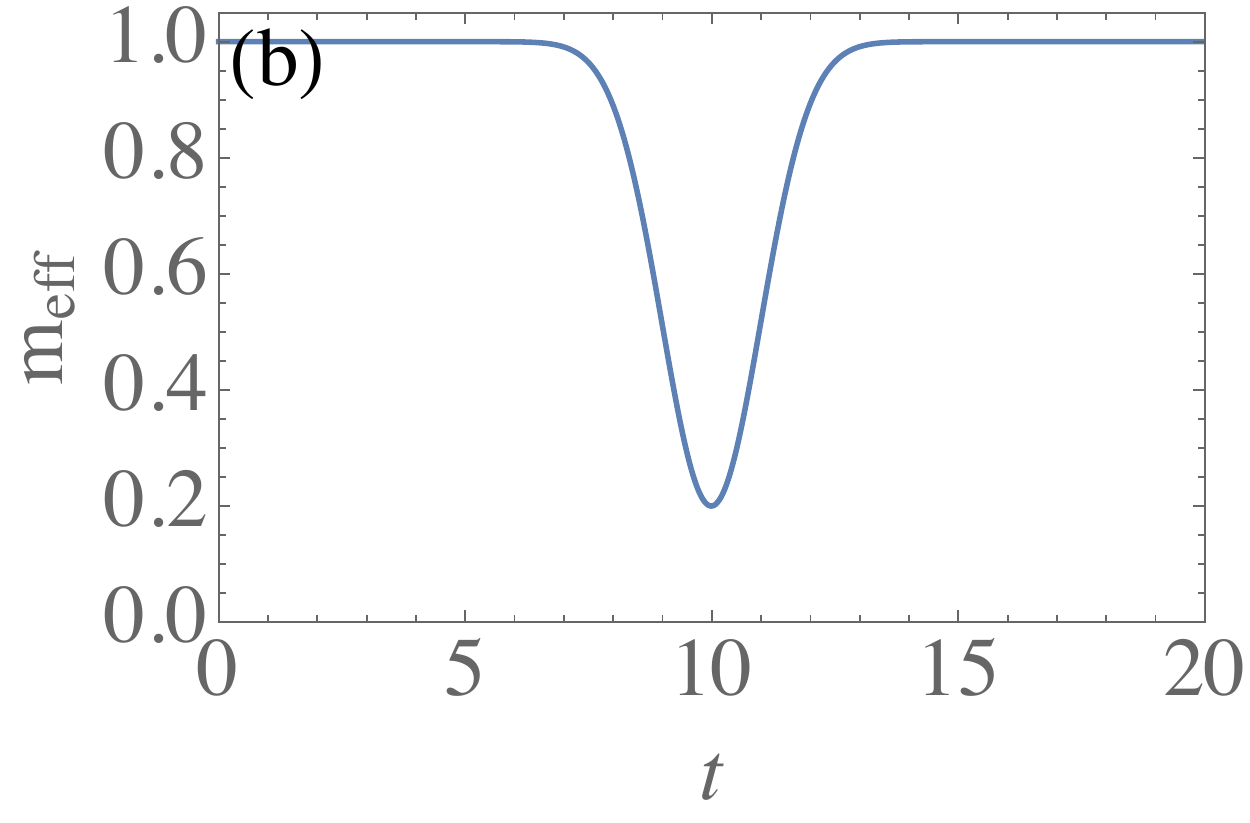}
\caption{Left, mean position of an initial positive-energy state in curved spacetime. Right, time profile of the conformal factor.}
\label{curved1}
\end{center}
\end{figure}

It turns out that physically interesting scenarios such as expanding spacetime (anti-de Sitter spacetime) produce only tiny effects that are unobservable, so instead we employ the `inverted square hat' profile of $m_{eff}(t)$ already introduced earlier in Sect.~\ref{sect3b}. Actually, we will use the smoothed version of this, by replacing the `inverted square hat' by an inverted Gaussian profile. Evidence of particle creation in this scenario is evident in Fig.~\ref{curved1}(a) where the induction of ZB by the excursion of $m_{eff}$ from it asymptotic value 1, is clearly visible. The initial state was constructed by superposing $\psi_-^k(x,0)$ from Eq.~(\ref{eestates}) with a Gaussian weight $\propto \exp[-4(k-0.1)^2]$. 

\section{Simulation of the Dirac equation in binary waveguide arrays}
In this section, we show how to simulate the 2 dimensional Dirac equation in binary waveguide arrays. We start off with a description of the coupled-mode theory of light propagation in a waveguide array and then demonstrate an equivalence between the discretised Dirac equation and the coupled-mode equations for a binary waveguide array. In particular, we show that the Dirac equation in 2 dimensional FRW spacetime can be straightforwardly implemented, paving the way towards experimental demonstration of the single particle analog of gravitational pair creation.

\subsection{Coupled-mode theory of waveguide arrays}
Propagation of an electromagnetic field $E$ in a medium is described by
\begin{equation}
\nabla^2 E - \frac{n^2}{c^2} \frac{\partial^2 E}{\partial t^2} = 0,
\end{equation}
where $n$ is the refractive index of the medium, spatially varying in general.
For a monochromatic field $E$ propagating predominantly in the $z-$direction, i.e., 
\begin{equation}
E= Re[E_0({\bf r})e^{i(kz-wt)}],
\end{equation}
one can make the so-called paraxial approximation which amounts to assuming $|\nabla^2E_0| \ll |k_z \partial_z E_0|$. In this case, the wave equation becomes the paraxial Helmholtz equation
\begin{equation}
k \frac{\partial E_0}{\partial z}  + \frac{i}{2}\left( k^2 -  \omega^2 \frac{n^2(r)}{c^2} \right) E_0 - \frac{i}{2} \nabla^2_{\perp} E_0     =  0,
\end{equation}
$\nabla^2_\perp$ denoting the Laplacian in $x$ and $y$ directions.
Let us choose $\omega/k = v_0 \equiv c/n_0$, where $v_0$ is the mean velocity in the medium and $n_0$ the mean refractive index. Then the above equation becomes
\begin{align}
2ik\partial_z E_0 + \nabla_\perp^2E_0-\frac{2k^2}{n_0}\Delta n E_0 = 0, 
\end{align}
where we have assumed that $n_0 \approx n$ and $\Delta n\equiv n_0 - n$. Lastly, define the reduced wavelength $\lambdabar = n_0/k$ to rewrite the equation as
\begin{align}
i\lambdabar \partial_z E_0 = -\frac{\lambdabar^2}{2n_0}\nabla^2_\perp  E_0 + \Delta n E_0.
\end{align}
Written this way, resemblance to the Schr\"odinger equation is obvious, the role of time being played by the spatial dimension $z$.

Now imagine periodically modulating the index of refraction in the plane perpendicular to the z-direction. An EM field is attracted to regions of increased index of refraction and mainly stays in the vicinity of these regions during propagation. The field, however, is not completely confined to these regions but leaks into the area between the waveguides, with evanescent tails. From the similarity with the Schr\"odinger equation, it is clear that one can apply the tight-binding approximation to describe the propagation of the EM field in this case. In optics, this is called the coupled mode approximation (see \cite{Szameit10}, which we are closely following). In a one dimensional lattice we get for the amplitude $c_n$ in the $n$th waveguide:
\begin{equation}
i\frac{\partial c_n}{\partial z}= k_n \left( c_{n+1} + c_{n-1} \right),
\end{equation}
where $k_n$ is the coupling strength, determined by the overlap between the transverse components of the modes in adjacent guides. Introducing yet another modulation such that alternating lattice sites have deep and shallow `potentials', one obtains the following coupled mode equations
\begin{equation}
\label{cmeqn}
i\frac{\partial c_n}{\partial z}= k_n \left( c_{n+1} + c_{n-1} \right) + (-1)^n \sigma c_n.
\end{equation}
In the next section we show that this equation is equivalent to the discretised Dirac equation.

\subsection{Discretising the Dirac equation}

\subsubsection{Flat spacetime}
To simulate the Dirac equation in flat 2D spacetime, 
\begin{equation}
i\frac{\partial \psi}{\partial t}=\left[-i\sigma_x{\partial_x}+m\sigma_z\right]\psi,
\end{equation}
in a waveguide array, we need to first change the differential equation to a difference equation by discretising the spatial coordinates. Choosing a discretisation length $d$, we define
\begin{align}
 \psi(x,t)  &\rightarrow  \psi(nd,t) \equiv \tilde{\psi}(n,t), \nonumber \\
{\partial_x}\psi_1(x,t) &\rightarrow  \frac{ \tilde{\psi}_1(n,t)-\tilde{\psi}_1(n-1,t)}{d}, \nonumber \\
{\partial_x}\psi_2(x,t) &\rightarrow  \frac{ \tilde{\psi}_2(n+1,t)-\tilde{\psi}_2(n,t)}{d},
\end{align}
where $n \in \mathbb{Z}$. 

\begin{figure}[ht]
\begin{center}
\includegraphics[width=0.8\columnwidth]{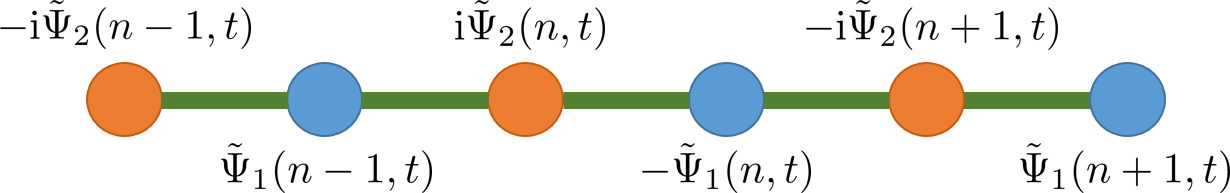}
\caption{Schematic illustration of how spinor components are assigned to waveguides. Different colours signify alternating refractive indices.}
\label{assignspinorcomp}
\end{center}
\end{figure}

To put this in the form of coupled mode equations, let us first define $\tc_{2n}(t) \equiv \tilde{\psi}_1(n,t)$ and $\tc_{2n-1} \equiv \tilde{\psi}_2(n,t)$. Then the Dirac equation becomes
\begin{align}
i\dot{\tc}_{2n}(t) = -\frac{i}{d}\left( \tc_{2n+1}(t) - \tc_{2n-1}(t) \right) + m\tc_{2n}(t), \nonumber \\
i\dot{\tc}_{2n-1}(t) = -\frac{i}{d}\left( \tc_{2n}(t) - \tc_{2n-2}(t) \right) - m\tc_{2n-1}(t).
\end{align}
To change the difference into the sum, we further define $c_{2n} = (-1)^n\tc_{2n} = (-1)^n\tilde{\psi}_1(n,t)$ and $c_{2n-1} = -i(-1)^n\tc_{2n-1} = -i(-1)^n\tilde{\psi}_2(n,t)$. The above equation changes to
\begin{align}
i\dot{c}_{2n}(t) = \frac{1}{d}\left( c_{2n+1}(t) + c_{2n-1}(t) \right) + mc_{2n}(t), \nonumber \\
i\dot{c}_{2n-1}(t) = \frac{1}{d}\left( c_{2n}(t) + c_{2n-2}(t) \right) - mc_{2n-1}(t),
\end{align}
which can be combined to
\begin{equation}
 i\dot{c_l}=-\frac{1}{d}(c_{l-1}+c_{l+1})+(-1)^{l} m c_l.
\end{equation}
Figure \ref{assignspinorcomp} provides a schematic illustration of how the spinor components are assigned to waveguides. 

The above equation is equivalent to Eq.~(\ref{cmeqn}) given that $k_n = 1/d$ and $\sigma = m$, as first noted by Longhi in his proposal to simulate zitterbewegung and Klein's paradox \cite{Longhi10a, Longhi10b}. Subsequently, these effects, which are predicted by the single-particle Dirac equation, have been observed in laser-written waveguide arrays \cite{Dreisow10,Dreisow12}.

\subsubsection{Curved spacetime}
To facilitate the simulation of the Dirac equation in 2 dimensional curved spacetime, we need to discretise Eq.~(\ref{curvedDirac}). In an optical simulation of single particle physics, the overall factor $\Omega$ is irrelevant, which means that we have at hand the Dirac equation with a spacetime-dependent mass as already noted in Sect.~\ref{sect:curvedDirac}. The generalisation of the coupled-mode equations Eq.~(\ref{cmeqn}) is then simple:
\begin{align}
i\frac{\partial c_n}{\partial z}= k_n \left( c_{n+1} + c_{n-1} \right) + (-1)^n \sigma_n (z) c_n.
\end{align}

\subsection{Optical simulation}
\begin{figure}[!htbp]
\begin{center}
\includegraphics[width=0.4\columnwidth]{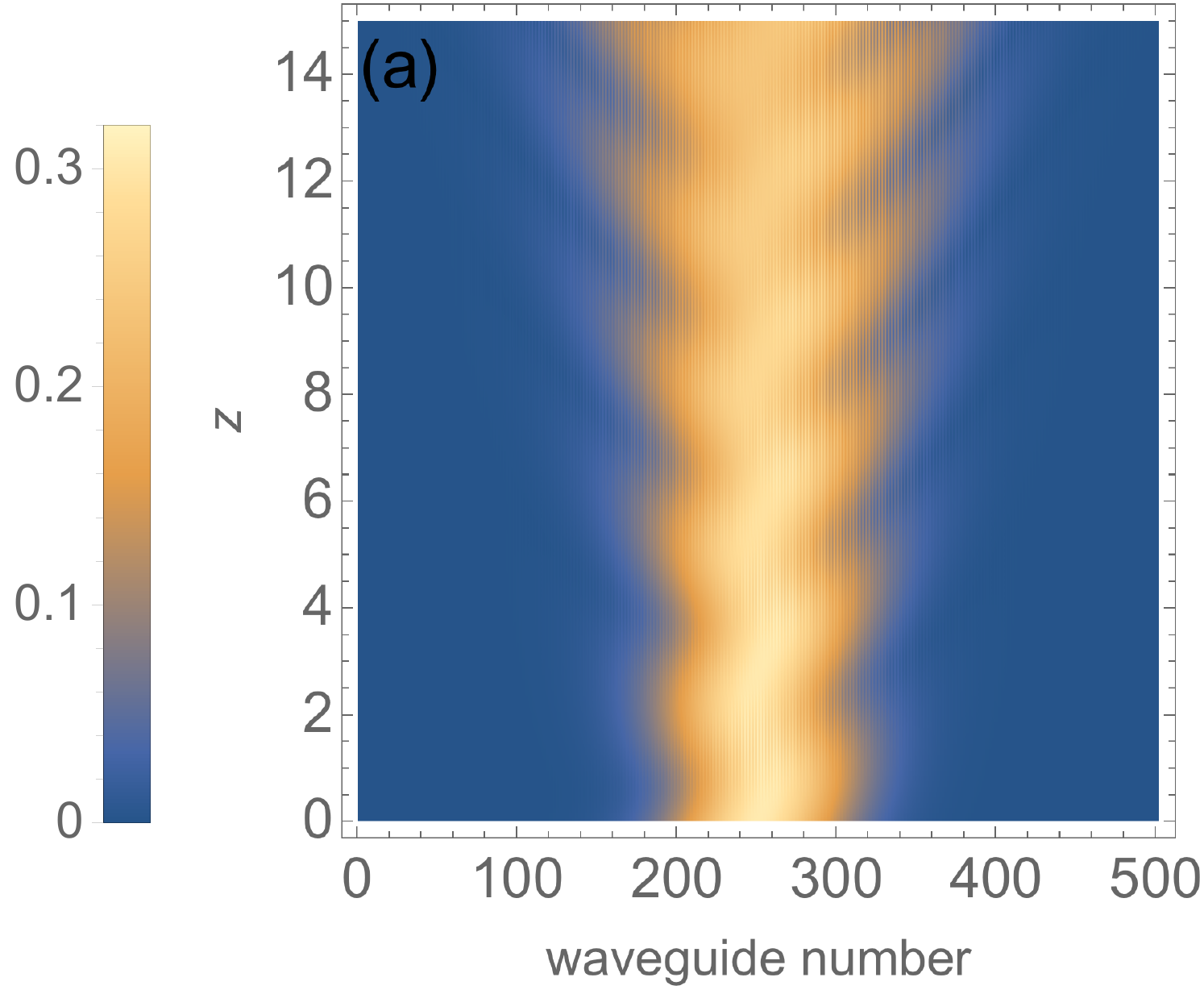}
\includegraphics[width=0.4\columnwidth]{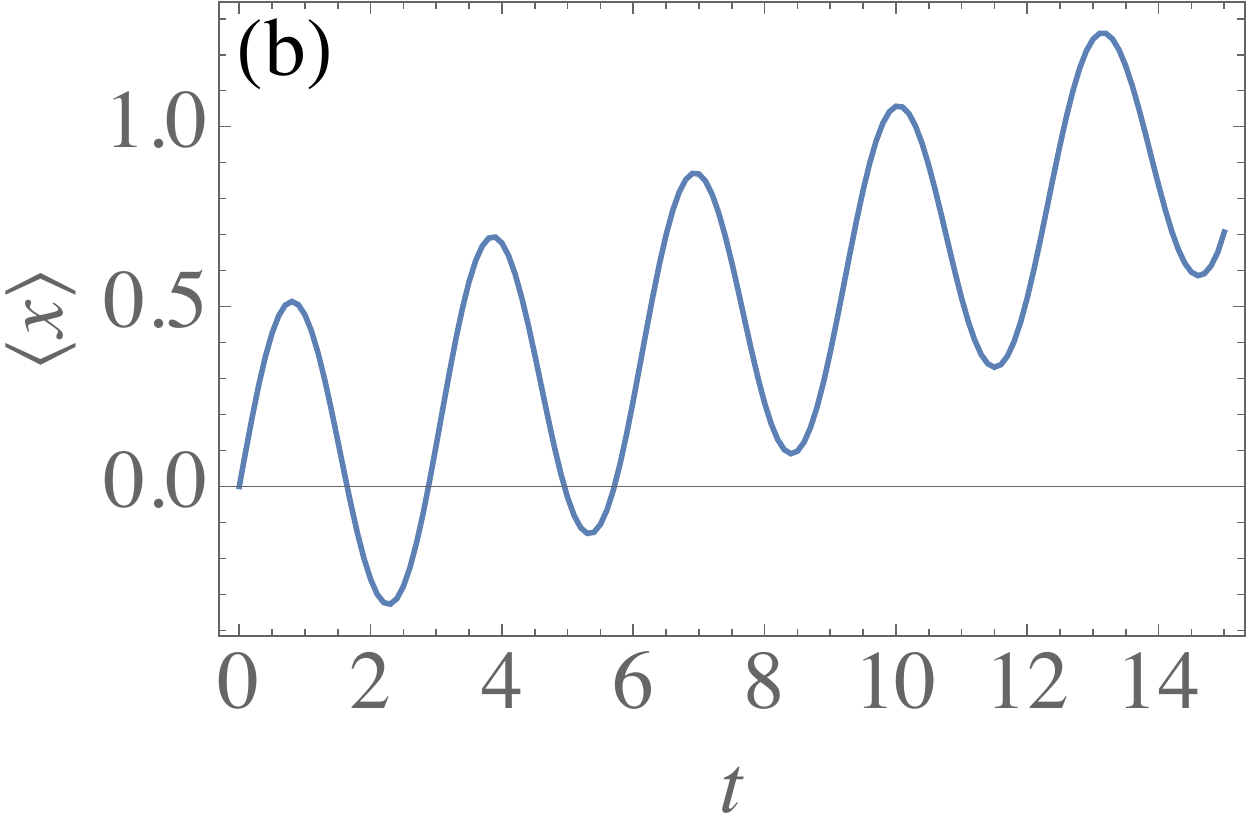}

\includegraphics[width=0.4\columnwidth]{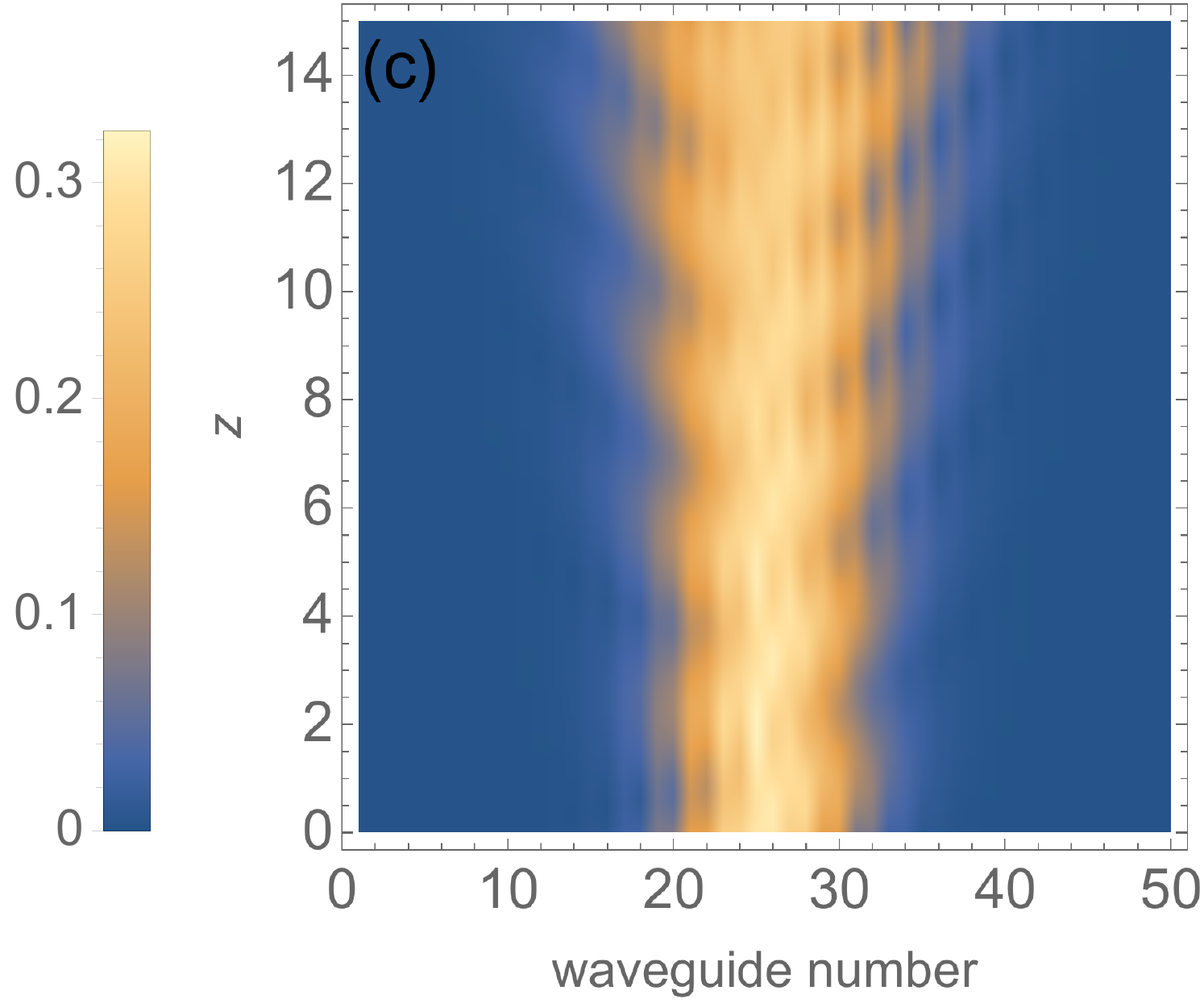}
\includegraphics[width=0.4\columnwidth]{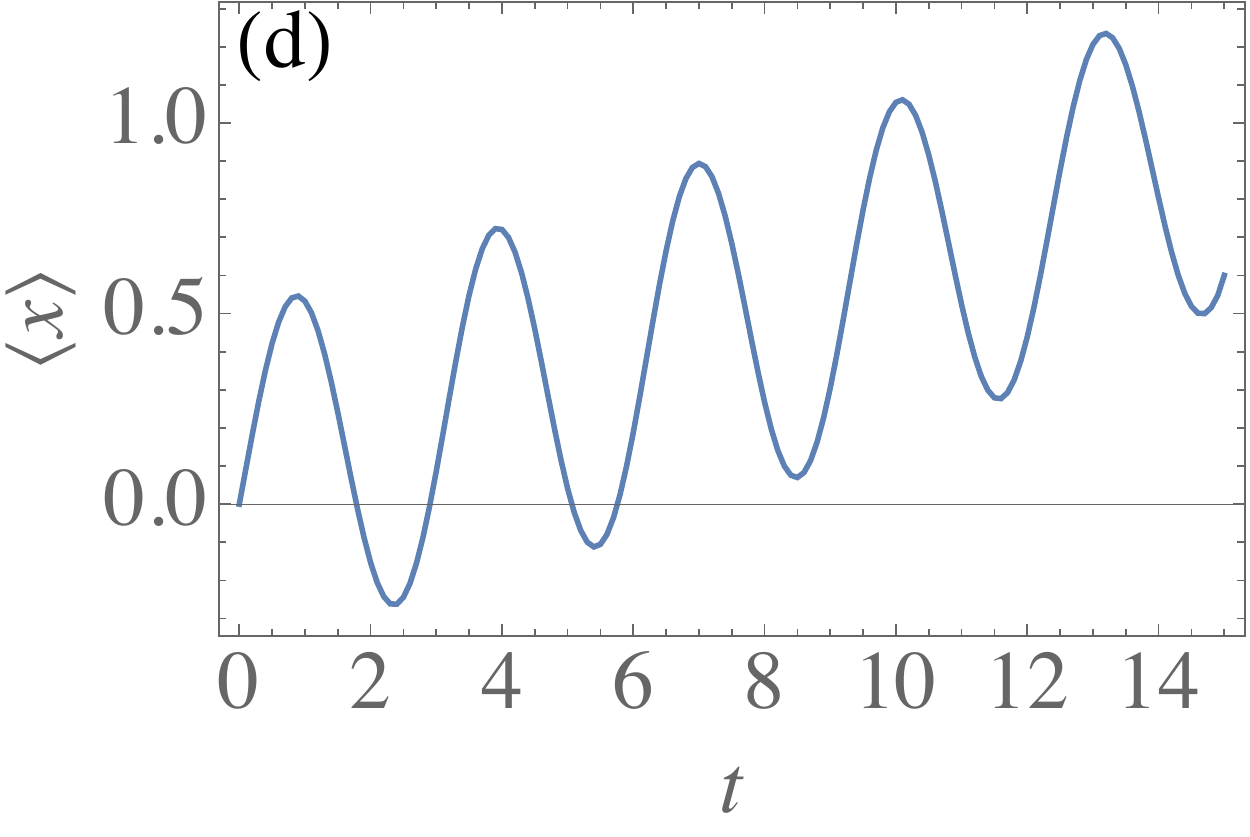}
\caption{Optical simulation of the Dirac equation in flat 2 dimensional spacetime. a) Absolute value of the mode amplitude, and b) average position of the simulated wave function as a function of time, for 502 waveguides. c) and d), analogous quantities for 50 coupled waveguides showing feasibility of the simulation despite strong discretisation. The initial state is $\propto \exp[-x^2/2\sigma^2-ik_0x](1,1)^T$ with the width $\sigma$ = 3, $k_0 = 0.1$, and mass $m = 1$. }
\label{simflat1}
\end{center}
\end{figure}

Let us start with flat spacetime. Figure \ref{simflat1} shows the evolution of a Gaussian spinor wave packet $\propto \exp[-x^2/18](1,1)^T$ when $m=1$. Simulation results for $N=502$ (requiring $k \approx 6.2$) waveguides are displayed in Fig.~\ref{simflat1}(a) and (b). They are in excellent agreement with the results in Sect.~\ref{dynamics}. However 502 is quite a large number and experiments are usually implemented with a much smaller number. To show the effects of discretisation we depict analogous results for $N=50$ ($k \approx 0.63$) in Fig.~\ref{simflat1}(c) and (d). Apart from coarse graining effects in visualisation, the wave packet evolution is seen to be remarkably accurate as exemplified by the average position.

\begin{figure}[!htbp]
\begin{center}
\vspace{0.5cm}
\includegraphics[width=0.4\columnwidth]{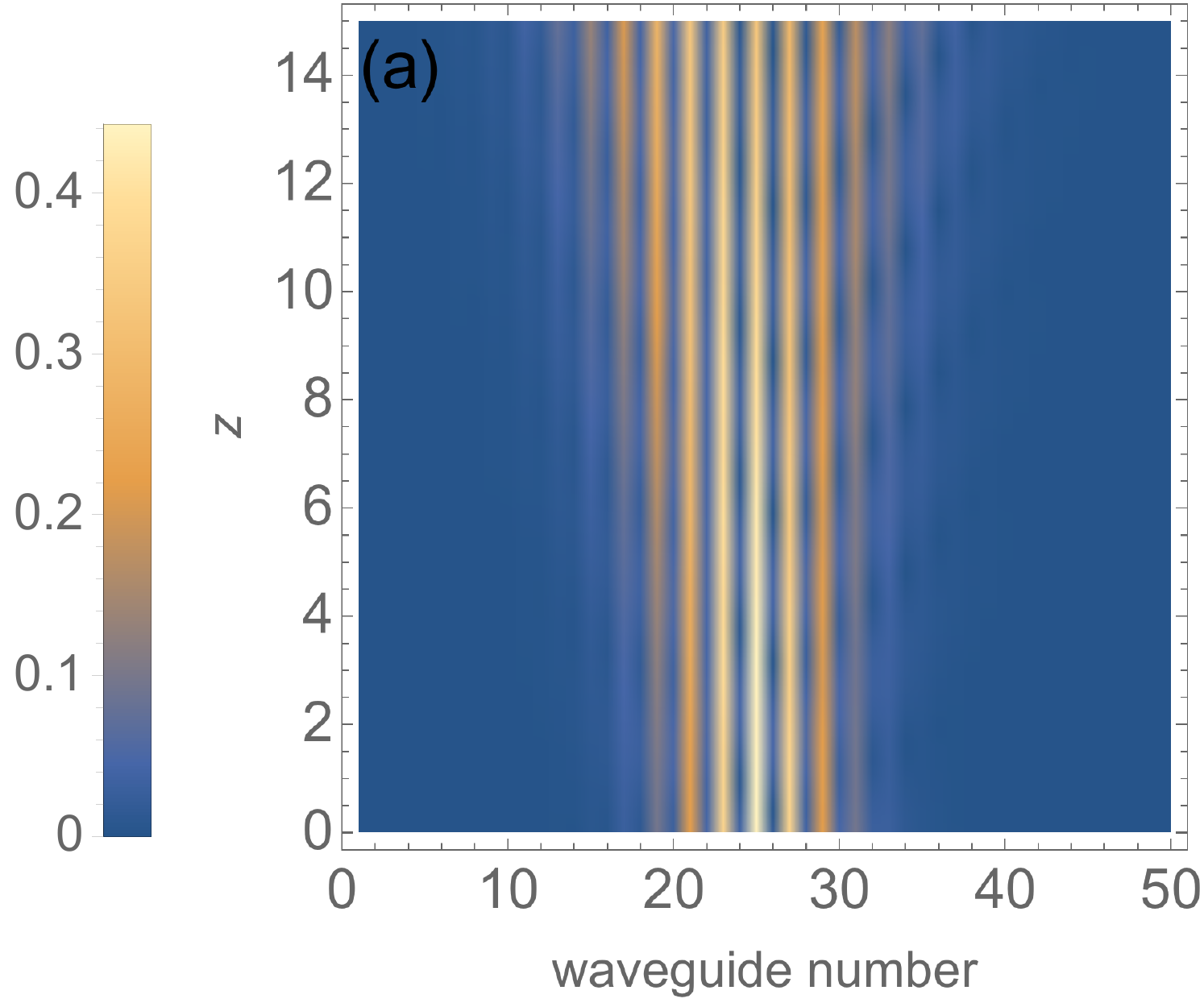}
\includegraphics[width=0.4\columnwidth]{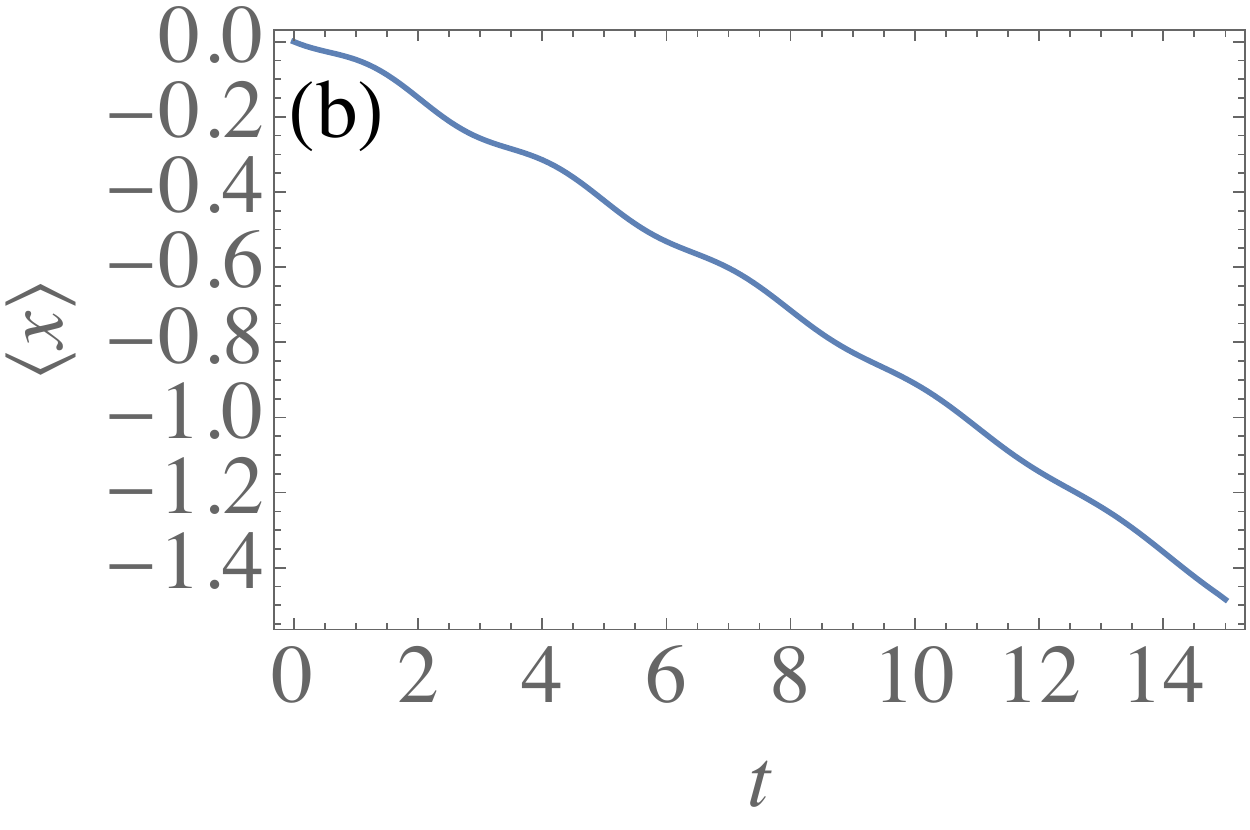}
\caption{Evolution of a negative energy spinor in flat spacetime. a) Absolute value of the mode amplitude, and b) average position of the simulated wave function as a function of time, for 50 waveguides. The initial state is $\propto \int dk \exp[-(k-k_0)^2/2\sigma_k^2] \phi_-^k$ with the width $\sigma_k = 1/(2\sqrt{2})$, $k_0 = 0.1$, and mass $m = 1$. }
\label{simflat2}
\end{center}
\end{figure}

Next, we simulate the conversion of a negative energy wave packet into a mixture of positive and negative energy wave packets. The latter is made by an arbitrary superposition of a negative energy eigen-spinor
\begin{align}
\phi_-^k = \frac{1}{\sqrt{2E_k(E_k + m)}}\begin{pmatrix} -k \\ E_k + m \end{pmatrix}.
\end{align}
In Fig.~\ref{simflat2}, we consider the evolution of a Gaussian-averaged spinor, $\propto \int dk \exp[-(k-k_0)^2/2\sigma_k^2] \phi_-^k$ with the width $\sigma_k = 1/(2\sqrt{2})$, $k_0 = 0.1$, and mass $m = 1$. There is a small amount of ZB as a result of discretisation (that goes away with the increasing number of waveguides), but the magnitude is quite small.

\begin{figure}[!htbp]
\begin{center}
\vspace{0.5cm}
\includegraphics[width=0.4\columnwidth]{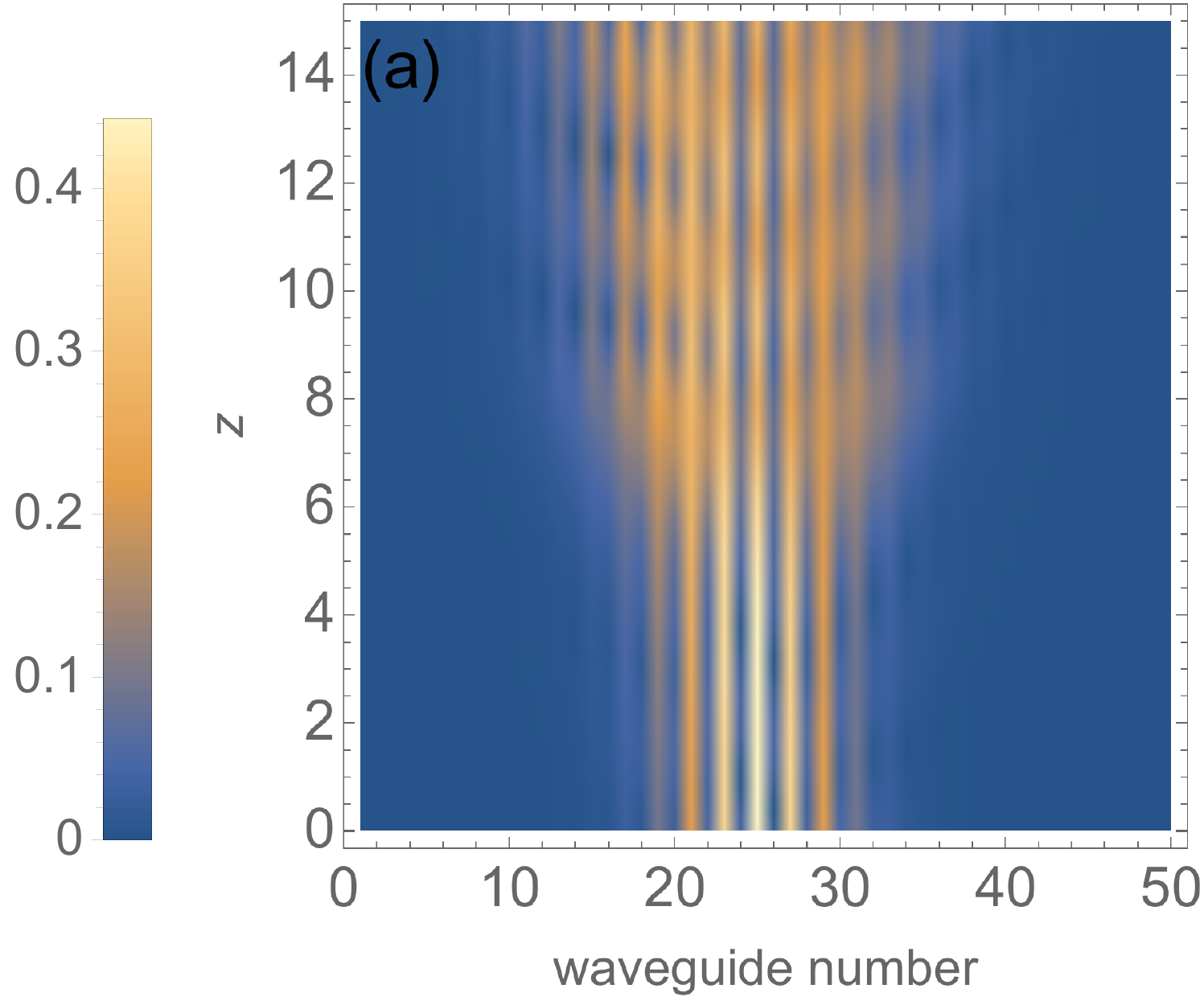}
\includegraphics[width=0.4\columnwidth]{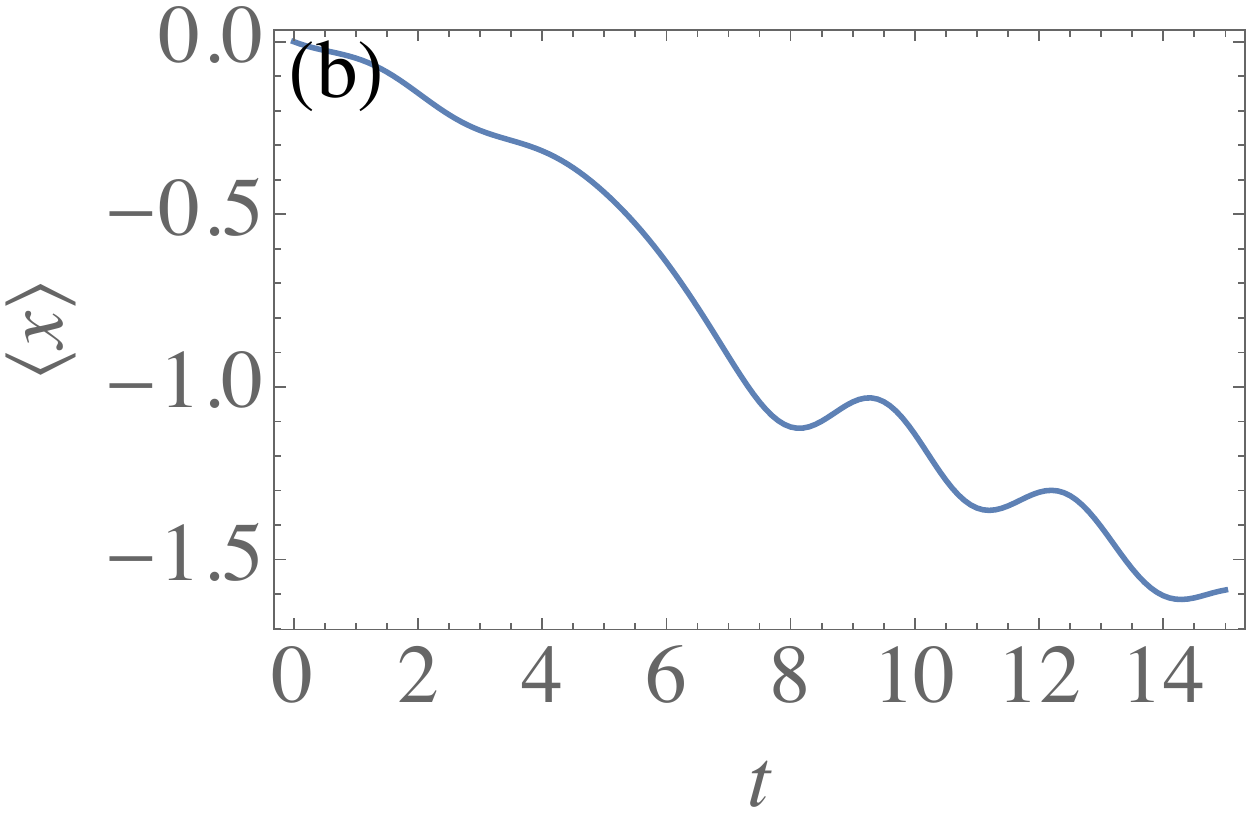}
\caption{Evolution of a negative energy spinor in FRW spacetime with inverted Gaussian conformal factor (cf Fig.~\ref{curved1}). a) Absolute value of the mode amplitude, and b) average position of the simulated wave function as a function of time, for 50 waveguides. The initial state is $\propto \int dk \exp[-(k-k_0)^2/2\sigma_k^2] \phi_-^k$ with the width $\sigma_k = 1/(2\sqrt{2})$, $k_0 = 0.1$, and mass $m = 1$. }
\label{simcurved}
\end{center}
\end{figure}

Finally, we show the evolution of a negative energy spinor in a FRW spacetime in Fig.~\ref{simcurved}, with the inverted Gaussian conformal factor as used in Fig.~\ref{curved1}. We see a good agreement with the exact numerical result shown in Fig.~\ref{curved1}, signifying the feasibility of optical simulation of particle creation.

\section{Conclusion}
We gave a pedagogical introduction to the Dirac equation in background curved spacetime and particle creation. Using the fact that the Dirac equation in the FRW metric is equivalent to the flat-spacetime Dirac equation with a time-dependent mass term, we demonstrated that a single-particle analog of particle creation can be observed in the dynamical evolution of spinor wave packets. In particular, we showed how a negative energy spinor gets mixed with a positive energy spinor when the conformal factor changes in time. Finally, we demonstrated that the Dirac equation in curved spacetime can be simulated in binary waveguide arrays, allowing direct experimental simulation of particle creation in curved spacetime. Although our example was for a time-dependent conformal factor, a general spacetime dependence can be easily simulated in waveguide arrays.

{\it Acknowledgments.}
D.G.A would like to acknowledge the financial support provided by the National Research Foundation and Ministry of Education Singapore (partly through the Tier 3 Grant ``Random numbers from quantum processes'' (MOE2012-T3-1-009)), and travel support by the EU IP-SIQS.

\bibliographystyle{unsrt}

\end{document}